\documentstyle[12pt]{article}
\newcommand{\beq}{\begin{equation}}
\newcommand{\eeq}{\end{equation}}

\def\half{{\textstyle{1\over2}}}

\def\p1half{{\textstyle{{{p+1}\over{2}}}}}

\def\23phalf{{\textstyle{{{23-p}\over{2}}}}}

\begin{document}
\thispagestyle{empty}
\begin{titlepage}

\bigskip
\hskip 3.7in{\vbox{\baselineskip12pt
\hbox{PSU-TH-243}\hbox{hep-th/0105244}}}

\bigskip\bigskip\bigskip\bigskip
\centerline{\large\bf
Finite Temperature Closed Superstring Theory:}

\medskip
\centerline{\large\bf
Infrared Stability and a Minimum Temperature}

\bigskip\bigskip
\bigskip\bigskip
\centerline{\bf Shyamoli Chaudhuri
\footnote{shyamoli@phys.psu.edu}
}
\centerline{Physics Department}
\centerline{Penn State University}
\centerline{University Park, PA 16802}
\date{\today}

\bigskip\bigskip
\begin{abstract}
We find that the gas of IIA strings 
undergoes a phase transition into a gas of IIB strings at the
self-dual temperature. A gas of free heterotic strings 
undergoes a Kosterlitz-Thouless duality transition 
with positive free energy and positive specific heat but vanishing
internal energy at criticality. We 
examine the consequences of requiring a tachyon-free thermal string 
spectrum. We show that in the absence of Ramond-Ramond fluxes the 
IIA and IIB string ensembles are thermodynamically ill-defined. The 10D
heterotic superstrings have 
nonabelian gauge fields and in the presence of a temperature dependent 
Wilson line background are found to share a stable and tachyon-free 
ground state at all temperatures starting from zero with gauge group 
$SO(16)$$\times$$SO(16)$. The internal energy of the heterotic string 
is a monotonically increasing function of temperature with a stable 
and supersymmetric zero temperature limit. Our results point to the 
necessity of gauge fields in a viable weakly coupled superstring
theory. Note Added (Sep 2005).
\end{abstract}

\end{titlepage}

\section{Introduction} 

It is generally assumed to be the case \cite{kirzhlinde,gpy} 
that in a quantum field theory known to be perturbatively renormalizable 
at zero temperature, the new infrared divergences introduced by a small 
variation in the background temperature can be self-consistently 
regulated by a suitable 
extension of the renormalization conditions, at the cost of introducing a 
finite number of additional counter-terms. The zero temperature 
renormalization conditions on 1PI Greens functions are conveniently applied 
at zero momentum, or at fixed spacelike momentum in the case of massless 
fields. Consider a theory with one or more scalar fields. Then the 
renormalization conditions 
must be supplemented by stability constraints on 
the finite temperature effective potential \cite{kirzhlinde}:
\begin{equation}
{{\partial V_{\rm eff.}(\phi_{\rm cl.})}\over{\partial 
\phi_{\rm cl.}}} = 0, \quad
\quad
{{\partial^2 V_{\rm eff.}(\phi_{\rm cl. })}\over{\partial \phi_{\rm cl.}^2 }} \equiv
G_{\rm \phi}^{-1} (k)|_{k=0}
\geq 0 \quad .
\label{eq:stab}
\end{equation}
Here, $V_{\rm eff.}(\phi_{\rm cl.})$$=$$\Gamma(\phi_{\rm cl.})/V\beta$, where
$\Gamma$ is the effective action functional, or sum of 
connected 1PI vacuum diagrams at finite temperature. In the absence of
nonlinear field configurations, and for perturbation theory in a small coupling 
about the free field vacuum, $|0>$, $\phi_{\rm cl.}$ is simply the 
expectation value of the scalar field: $\phi_{\rm cl.}$$=$$<0 | \phi(x) | 0 >$. 
The first condition holds in the absence of an external source at every 
extremum of the effective potential. 
The second condition states that the renormalized masses of physical fields 
must not be driven to imaginary values at any non-pathological and stable 
minimum of the effective potential.

\vskip 0.1in
The conditions in Eq.\ (\ref{eq:stab}) are rather familiar to string theorists.
Weakly coupled superstring theories at zero temperature are replete with 
scalar fields and their vacuum expectation values, or moduli, parameterize a 
multi-dimensional space of degenerate vacua. Consider the effect of
an infinitesimal variation in the background temperature. Such an effect will 
necessarily break supersymmetry, and it is well-known that
in the presence of a small spontaneous breaking of supersymmetry the dilaton
potential will generically develop a runaway direction, signaling an instability
of precisely the kind forbidden by the conditions that must be met by a 
non-pathological ground state \cite{dinesei}. Namely, while the zero temperature 
effective potential is correctly minimized with respect to the renormalizable 
couplings in the potential and, in fact, vanishes in a spacetime supersymmetric 
ground state, one or more of the scalar masses 
is driven to imaginary values in the 
presence of an {\em infinitesimal} variation in background
temperature. Such a quantum
field theory is simply unacceptable both as a self-consistent effective field 
theory in the Wilsonian sense \cite{rg}, and 
also as a phenomenological model for a 
physical system \cite{kirzhlinde}. The same conclusion must hold for a weakly
coupled superstring theory with these pathological properties. 

\vskip 0.1in
In the absence of gauge fields, the low temperature behavior of the
ten-dimensional type II superstrings is pathological in the 
sense described above \cite{aw,us}. While the zero 
temperature free string Fock space is supersymmetric and tachyon-free,
an equilibrium thermal ensemble of either closed superstring,
type IIA or type IIB, contains a tachyonic physical state at infinitesimal 
temperature \cite{aw}. It follows that, {\em in the absence of gauge 
fields, flat spacetime is an unstable ground state for the 
ten-dimensional weakly coupled type II string theories under a small 
variation in the background temperature.} This result, first pointed 
out by us in \cite{us}, can be stated within the framework of the 
renormalization conditions for a consistent finite 
temperature closed string perturbation theory: there are no non-tachyonic 
solutions to the renormalization conditions in the vicinity 
of the zero temperature ground state, in the absence of gauge 
fields that can potentially arise in a non-trivial Ramond-Ramond sector.

\vskip 0.1in
On the other hand, in the 
presence of temperature dependent Wilson lines, the 10D
heterotic superstrings have {\em a stable and tachyon--free 
finite temperature ground state at all temperatures starting from 
zero, but with gauge group $SO(16)$$\times$$SO(16)$.} 
The low energy effective theory is a finite temperature gauge-gravity theory, 
formulated in modified axial gauge with the Euclidean time component of the
vector potential, $A_0$, set equal to a temperature dependent constant. The
additional massless vector bosons that lead to an enhancement of the 
gauge group in the supersymmetric zero temperature limit to either 
$E_8$$\times$$E_8$, or $SO(32)$,
are massive at finite temperature. As 
was shown in \cite{us}, an analogous result holds for the finite
temperature type I string. As explained in the conclusions
\cite{us}, at least for infinitesimal temperatures,
it should be possible to interpret the type I thermal string
as the strong coupling dual of the thermal
$SO(16)$$\times$$SO(16)$ heterotic string. We defer further 
discussion of finite temperature open and closed string theory 
to \cite{type1}. 

\vskip 0.1in
In this paper we will compute the generating functional 
of connected one-loop vacuum string graphs in the embedding 
space $R^9$$\times$(a compact 
one-dimensional target space with Euclidean signature) for
each of the 10D fermionic closed string theories. As
pointed out in \cite{us,bosonic}, there is an ambiguity in
the Euclidean time prescription that becomes apparent when we 
try to apply it to finite temperature string theory: a 
one-dimensional compact space can have the topology of a 
circle or that of an interval, which assumption yields the 
correct answer for the thermal spectrum of a gas of free 
strings? In \cite{bosonic} we resolve this ambiguity by 
showing that a ${\rm Z}_2$ twist in Euclidean time is necessary 
for the absence of spurious massless vector bosons in the 
finite temperature spectrum. Given the residual $d$-dimensional 
gauge invariance, the number of propagating degrees of freedom 
in a finite temperature gauge theory in $d$ spatial dimensions 
correspond precisely to those in an axial gauge quantization, 
$A^0$ $=$ $0$, of the Euclidean $d$$+$$1$ dimensional gauge 
theory \cite{bernard}. Thus, by requiring that the massless 
states in the thermal string spectrum match with the physical 
degrees of freedom in a consistent finite temperature
gauge-gravity the ambiguity in the Euclidean time prescription
is resolved. 

\vskip 0.1in
In the fermionic strings: IIA, IIB, or heterotic, the 
${\rm Z}_2$ orbifold twist in Euclidean time naturally 
extends to a ${\rm Z}_2$ superorbifold twist, identifying 
an appropriate fixed line of ${\hat c}$ $=$ $1$ superconformal 
field theories parameterized by interval length, or inverse
temperature: $\beta$ $=$ $\pi r_{\rm circ}$. 
The super-orbifold action must be consistent with the 
appropriate superconformal invariance on the worldsheet.
As in the bosonic string \cite{us, bosonic}, we find that
the Hagedorn radius $(2\alpha^{\prime})^{1/2}$ of a gas of 
fermionic strings always coincides with the Kosterlitz-Thouless 
point on a fixed line of superorbifolds with interval
length $\pi \alpha^{\prime 1/2}$: the ${\hat c}$ $=$ $1$ 
orbifold of the circle compactified superconformal field 
theory at the self-dual radius. The nature of the duality
transition at the Kosterlitz-Thouless point differs:
the II A and IIB thermal strings are related by a T-duality 
transformation in Euclidean time, while the heterotic string 
is self-dual.

\vskip 0.1in
Following the approach taken by Polchinski \cite{poltorus},
our starting point for a discussion of string thermodynamics
is the world-sheet path integral representation of the
generating functional of connected one-loop vacuum string graphs:
$W(\beta)$ $\equiv$ ${\rm ln}$ ${\rm Z}(\beta)$. But, as explained
above \cite{us,bosonic}, we introduce
a ${\rm Z}_2$ twist in Euclidean time in order to match with the
physical degrees of freedom in the low energy finite temperature
gauge-gravity theory:
$W(\beta)$ is computed in the embedding space
$R^9$$\times$$S^1/Z_2$ with interval length identified as
the inverse temperature $\beta$. The thermodynamic relations for the
free string gas take the form:
\begin{equation}
W(\beta) \equiv {\rm ln}~ Z , \quad
F(\beta) = -W/\beta , \quad
\rho(\beta) = -W/V \beta , \quad
      U(\beta) = - T^2 \left ( {{\partial W}\over{\partial T}} \right )_V =
             \left ( {{\partial W}\over{\partial \beta}} \right )_V
\quad .
\label{eq:freene}
\end{equation}
$F$ is the Helmholtz free energy of the ensemble of free strings,
$U$ is the internal energy, and $\rho$ is the finite temperature
effective potential. Notice that $\rho$ is the finite temperature analog of
the one-loop vacuum energy density or cosmological constant, $\rho_0$.
$\beta$ is a continuously-varying background parameter describing a 
one-parameter family of consistent ground states
of string theory, characterized by the $\beta$ dependent effective potential.
In practise, we will find it impossible to satisfy the requirement of 
a tachyon-free spectrum over the entire temperature range in the case
of the type II strings, at least in the absence of a Ramond-Ramond 
sector. For the heterotic string, the presence of gauge fields permits
a one-parameter family of tachyon-free ground states over the entire
temperature axis upon inclusion of a temperature dependent Wilson line.
As in the bosonic string \cite{bosonic}, the thermodynamic potentials 
of the free string ensemble are obtained directly from the generating
functional of connected vacuum graphs: ${\rm ln}$ ${\rm Z}(\beta)$, as
opposed to ${\rm Z}(\beta)$. Thus, we never need address the troubling
issue of defining the thermodynamic limit of the canonical ensemble, 
since we never compute the canonical partition function for string states 
directly in the path integral framework. 

\vskip 0.1in
An important new issue is the determination of modular 
invariant but supersymmetry breaking temperature dependent
phases in the fermionic closed string path integral. 
We will find that the type II string theories 
display an unusual phenomenon: unlike the bosonic
string, a stable and tachyon-free thermal 
ensemble exists only beyond a string scale 
minimum temperature. The minimum temperature
is the temperature at which the leading tachyonic 
momentum mode turns massless.
Thermal duality interchanges the momentum and winding modes, respectively,
of IIA and IIB string theories. Consequently, each type II string theory
also exhibits a characteristic maximum temperature,
at which the first winding mode turns tachyonic. 
The necessity for both winding and momentum modes in either
type II string follows from modular invariance.
Finally, the inclusion of a duality invariant phase ensures that
upon mapping IIA winding modes to IIB momentum modes, and vice versa,
the expression for the vacuum functional of the IIA theory is 
likewise mapped to that for the IIB theory.
Thus, in the absence of background fields, either type II free 
string ensemble is stable, but only within
a string scale temperature regime containing the self-dual 
point, $T_C$$=$$1/\pi \alpha^{\prime 1/2}$. {\em The
Kosterlitz-Thouless phase transition at $T_C$
should now be viewed as a continuous phase transition mapping 
the thermal IIA string to the thermal IIB string.}
A similar result holds for the type I string theory
and its thermal dual, the type ${\tilde{{\rm I}}}$ 
string \cite{us,type1}.

\vskip 0.1in
The 10D $E_8$$\times$$E_8$ and $SO(32)$ heterotic strings share
a common finite temperature ground state, and the vacuum functional of 
this nonsupersymmetric and tachyon-free string theory is 
self-dual under thermal duality transformations. 
For either 10D heterotic superstring, we will show that 
turning on a temperature dependent Wilson line background permits a
stable and tachyon-free nonsupersymmetric ground state at all 
temperatures starting from zero, but with non-abelian gauge group
$SO(16)$$\times$$SO(16)$. The expressions for the Helmholtz and 
Gibbs free energies are finite at the Hagedorn temperature, 
confirming the absence of any exponential divergences in the 
free energy of a tachyon-free string theory and, consequently, 
the absence of a Hagedorn phase transition as anticipated 
in \cite{us,bosonic}.

\vskip 0.1in
The self-dual heterotic string gas displays the 
Kosterlitz-Thouless duality transition described in section 4
of \cite{bosonic}. The Helmholtz and Gibbs free energies are
minimized at criticality, and the internal energy vanishes.
As explained in \cite{bosonic}, the cancelation is simply 
understood as a shift in balance between energy and entropy:
winding and momentum modes contribute equally to both the 
free energy and internal energy at criticality, and the gas
of free strings transitions from a phase of bound vortex
pairs to the high temperature long string phase dominated
by winding modes \cite{kosterlitz}. Similar arguments 
explain the zeroes of the one-loop effective potential 
found in the numerical computations of Ginsparg and Vafa 
\cite{ginine,gv}, and of Itoyama and Taylor \cite{itoyama}. 

\vskip 0.1in
The free energy of the gas of $SO(16)$$\times$$SO(16)$
heterotic strings is positive at criticality, a consequence
of an excess of spacetime fermions over spacetime bosons 
in the physical state spectrum. As in the bosonic string 
\cite{poltorus,bosonic}, 
this aspect of the $R^9$$\times$$S^1/{\rm Z}_2$ Euclidean 
time ground state simply mirrors an analogous result for the
corresponding Lorentzian signature, nonsupersymmetric and 
tachyon-free, 10D heterotic string ground state found in 
\cite{sw,agmv,dh,klt}. As noted in both \cite{agmv} and 
\cite{dh}, unlike the tachyonic ground states, the 10D 
nonsupersymmetric and tachyon-free $O(16)$$\times$$O(16)$ 
heterotic ground state has {\em positive} cosmological 
constant. As anticipated in \cite{bosonic}, we can verify
that the Gibbs and Helmholtz free energies are minimized 
at the self-dual temperature, and the vanishing internal 
energy results in {\em positive} specific heat at the
transition temperature. This is an encouraging indication 
of the thermodynamic stability of the finite 
temperature ground state.

\vskip 0.1in
The plan of this paper is as follows. In section 2, 
we begin with a brief overview of the stability conditions 
at finite temperature on $W(\beta)$, the generating functional
of connected vacuum graphs in string theory.
This is followed by a discussion of the implementation
of thermal boundary conditions on fermions in the superstring path 
integral in section
2.1. The Polyakov path integral representation 
of the string vacuum functional for an equilibrium ensemble of
type IIA or type IIB strings at finite temperature, including 
determination of the supersymmetry breaking phases, is
presented in Section 3. In Section 3.1, we obtain 
the minimum temperature for tachyon-free IIA or IIB 
thermal ensembles, at which the leading tachyonic momentum 
mode turns massless. We clarify the existence of a continuous
Kosterlitz-Thouless phase transition between the thermal type IIA
and thermal type IIB strings at the self-dual Hagedorn point.
The source of the tachyonic instability at $T_{\rm min}$, and 
its possible resolution, are discussed briefly.

\vskip 0.1in
Section 4 contains a description of the nonsupersymmetric
9D $SO(16)$$\times$$SO(16)$ heterotic string, the
tachyon-free finite temperature ground state common to both 10D
heterotic superstrings. 
We explain how the 9D $SO(16)$$\times$$SO(16)$ heterotic
string is obtained by turning on a temperature-dependent
timelike Wilson line in either zero temperature
supersymmetric 10D heterotic string \cite{us}. An infinitesimal
variation in temperature away from the zero temperature limit 
of the heterotic superstring results in the breaking of both 
the supersymmetry, and a partial breaking of the
non-abelian gauge symmetry, with inverse
temperature playing the role of a small control parameter.
The mechanism for symmetry breaking in the vicinity of the
zero temperature ground state appears to have no field theoretic
correspondence that I am aware of, but it can be 
seen to be a consequence of requiring a modular invariant 
and tachyon-free thermal string spectrum. 
From the viewpoint of the low-energy, effective, 
finite temperature gauge-gravity theory,
the timelike Wilson line background is interpreted 
as the imposition of a particular axial gauge condition. 
We explain why the twist II orbifold
construction of \cite{itoyama} suggests the reverse, namely,
the spontaneous restoration of symmetries at high temperature.
For comparison, we also clarify 
the finite temperature behavior of the
{\em non-supersymmetric} and tachyon-free 10D $SO(16)$$\times$$SO(16)$
heterotic string theory, first discovered in 
\cite{sw,agmv,dh,klt}. 
We conclude with a brief discussion of the Helmholtz free 
energy and, more generally, of the hierarchy of thermal duality 
relations satisfied by the thermodynamic potentials of the 
free heterotic string ensemble. The conclusions explain the 
implications of
these results, both for the further elucidation of string 
thermodynamics, and also for the clarified understanding 
of dualities in the moduli space of String/M theory. 
We discuss some open questions for future work.

\section{Thermal Fermionic String Theories}

In this section, we will compute the generating functional of connected 
one-loop vacuum string graphs in the Euclidean embedding 
space $R^9$$\times$$S^1/Z_2$ for both the type II and heterotic closed 
strings. As was shown in \cite{bosonic}, the expression for the
string vacuum functional is the starting point for a 
discussion of string thermodynamics. The Helmholtz free energy and
remaining thermodynamic potentials are obtained by simply taking
appropriate derivatives with respect to inverse temperature, or volume, 
using the usual thermodynamic definitions for a canonical ensemble. 

\vskip 0.1in
As in finite temperature field theory \cite{kirzhlinde}, a small
variation in the background temperature can generically 
result in tachyonic instabilities in 
the thermal free string ensemble if the mass of a physical, or unphysical, 
state in the mass spectrum is driven imaginary \cite{poltorus,aw,tan}. 
It was pointed out in \cite{us} that the generating functional of 
vacuum graphs in a fully consistent string theory at finite temperature--- 
heterotic, type I, type IIA or type IIB, should have {\em no infrared 
instabilities in the low
temperature regime}. In other words, while it is reasonable to allow for 
the possibility of a genuine phase transition where the Helmholtz 
free energy is divergent for all temperatures beyond some {\em finite 
transition temperature}, the zero temperature free string
Fock space of a consistent supersymmetric string theory should not contain
physical states whose masses are driven to imaginary values at {\em
infinitesimal temperature}. As mentioned in the introduction, in
the absence of background fields, all of
the weakly coupled ten-dimensional supersymmetric strings are 
pathological in this respect. 

\vskip 0.1in
As described in \cite{bosonic}, the starting point for a discussion
of string thermodynamics is $W$$\equiv$${\rm ln}Z$, 
the generating functional of connected
one-loop vacuum string graphs in the Euclidean embedding space 
$R^9$$\times$$S^1/Z_2$, where the interval in the imaginary time direction
has length equal to the inverse temperature $\beta$$=$$\pi r_{\rm circ.}$.
This yields the one-loop contribution to the finite temperature effective 
potential in string theory via the relation: 
$\Gamma_{\rm eff.}$$=$$ -W/\beta V$$=$$\rho(\beta)$ \cite{poltorus}.
Successive partial derivatives of $W$ with respect to $\beta$
at fixed spatial volume will yield the thermodynamic potentials:
\begin{equation}
F =  -W/\beta , \quad U = T^2 {{\partial}\over{\partial T}} W ,  \quad
P =  - \left ( {{\partial F }\over{\partial V}} \right )_T  ,
\quad
S =  - \left ( {{\partial F }\over{\partial T}} \right )_V  ,
\quad
C_V =  T \left ( {{\partial S}\over{\partial T}} \right )_V \quad ,
\label{eq:can}
\end{equation}
where $F$ and $U$ denote, respectively, the Helmholtz free energy
and the internal energy,
$S$ denotes the entropy, and $C_V$ denotes
the specific heat at constant volume, of the canonical ensemble.
For the gas of free strings, the 
finite temperature effective potential is the extensive
vacuum energy density: $\rho$$=$$F/V$ \cite{poltorus}.
It is easy to see that the pressure of the free string gas
vanishes, so that the enthalpy equals the internal
energy: $H$$=$$U$, and, consequently, that the Gibbs and Helmholtz
free energies are identical: $G$$=$$H$$-$$TS$$=$$U$$-$$TS$$=$$F$.
We emphasize that this holds in the absence of string interactions.

\vskip 0.1in
We will require that the finite temperature effective 
potential or, equivalently, the Helmholtz free energy,
are free from thermal
tachyons. Remarkably, for the heterotic string we will find
that a tachyon-free ensemble exists at all temperatures starting 
from zero, in the presence of a temperature dependent Wilson 
line background, a result that could have been anticipated from the
earlier works \cite{sw,agmv,dh,klt,ginine}.

\subsection{Thermal Boundary Conditions in Fermionic Strings}

In this subsection, we give a discussion of 
thermal boundary conditions and the precise action 
of the orbifold group on world-sheet fermions.
Recall that in the imaginary time formalism, spacetime fields obeying
Fermi statistics are to be expanded in a Fourier basis of antiperiodic 
modes alone. This implies antiperiodic boundary conditions for the 
spacetime fermionic modes of the string path integral in the imaginary 
time direction. The spacetime spin and statistics of fields 
in the Euclidean embedding space can be determined in string theory by 
suitably adapting the GSO prescription, given by a specification
of the world-sheet spin 
and statistics of every state in the free string spectrum 
\cite{sw,ckt,dh,agmv,klt,polchinskibook}. We will use the 
Ramond-Neveu-Schwarz (RNS) formulation to compute the spectrum of 
physical states in a fermionic string theory. Fermions on the worldsheet 
are simultaneously spacetime vectors and worldsheet spinors.

\vskip 0.1in
There is no spin-statistics theorem in two dimensions and we are
therefore free to choose arbitrary boundary conditions on the 
world-sheet fermions \cite{ckt}. We emphasize that such a generalized 
sum is necessitated by modular invariance and the thermal boundary 
conditions. In practise, it is often easier to arrive at the required
one-loop amplitude by manipulating modular invariant blocks
of Jacobi theta functions which preserve spacetime Lorentz invariance, 
and consequently, the world-sheet superconformal invariances. Thus, 
we identify modular invariant blocks of eight spin structures pertaining 
to the eight transverse world-sheet fermions. These blocks are weighted 
by, a priori, undetermined temperature-dependent phases. The form of 
the temperature dependent phases is guided by requiring a modular 
invariant partition function which interpolates continuously between 
the supersymmetric zero temperature result, and a finite temperature
expression which transforms correctly under a thermal duality 
transformation.

\vskip 0.1in
To motivate our result, it is helpful to consider the generalization 
of the thermal closed bosonic string given in \cite{us,bosonic}. 
The action of a fermionic R-orbifold group on $\psi^0$, $ X^0$, 
consistent with the preservation of an N=1 world-sheet superconformal 
invariance, satisfying thermal boundary conditions, and transforming
correctly under a thermal duality transformation, remains to be specified. 
To begin with, we recall
the analysis in \cite{dgh} where the distinct possibilities for 
fixed lines of N=1 superconformal theories with superconformal central 
charge ${\hat c}$ $=$ ${{2}\over{3}} c$ $=$ $1$ were classified. We can 
identify points under the reflection $R$,
$ X^0 \simeq - X^0 $, with fundamental region the half-line,
$X^0 \geq 0$, supplementing with the periodic identification,
$t^w : $ $ X^0 \simeq X^0 + 2 \pi w r_{\rm circ.}$, $w \in {\rm Z}$.
The resulting fundamental region is the interval, 
$0 \leq X^0 < \pi r_{\rm circ.}$ \cite{polchinskibook}.
An N=1 world-sheet supersymmetry requires 
that $\psi^0 \simeq - \psi^0$ under $R$, and the periodic 
identification $t^w$ ordinarily preserves the 
superconformal generator, $\psi^0 \partial_z X^0$,
leaving the fermions invariant. The partition function of
an Ising fermion is the sum over spin structures (A,A), (A,P), 
(P,A), and (P,P), manifestly invariant under $\psi$ $\to$ $-\psi$
\cite{dgh}. Thus, two fixed lines of ${\hat c}$$=$$1$ theories are 
obtained by simply
tensoring the Ising partition function with either 
${\rm Z}_{\rm circ.}$, or ${\rm Z}_{\rm orb}$, from 
the previous section \cite{dgh}:
\begin{equation}
{\rm \hat Z}_{\rm circ.} = 
{\rm Z}_{\rm circ.} {\rm Z}_{\rm Ising}, \quad   
{\rm \hat Z}_{\rm orb.} = 
{\rm Z}_{\rm orb.} {\rm Z}_{\rm Ising} , \quad 
{\rm Z}_{\rm Ising} = 
\half \left ( |{{\Theta_3}\over{\eta}}| +
 |{{\Theta_4}\over{\eta}}| +
 |{{\Theta_2}\over{\eta}}| \right ) \quad .
\label{eq:is}
\end{equation}
Accompanying $R$, $t^w$ with
the action of the Z$_2$-moded generator,
$(-1)^{N_F}$, where $N_F$ is spacetime fermion number, one 
can define additional super-orbifold theories, ${\rm \hat Z}_{\rm sa}$,
${\rm \hat Z}_{\rm so}$, and ${\rm \hat Z}_{\rm orb.'}$, giving a 
total of five fixed lines of ${\hat c}$ $=$ $1$ theories \cite{dgh}:
\begin{equation}
{\rm \hat Z}_{\rm orb.'} = R (-1)^{N_F} {\rm \hat Z}_{\rm circ.} , \quad
{\rm \hat Z}_{\rm sa} =  (-1)^{N_F} e^{2\pi i \delta(p)} {\rm \hat Z}_{\rm circ.}, \quad
{\rm \hat Z}_{\rm so} = R {\rm \hat Z}_{\rm sa} 
= R (-1)^{N_F} e^{2\pi i \delta(p)} {\rm \hat Z}_{\rm circ.} \quad ,
\label{eq:aff}
\end{equation}
where $\delta(p)$ acts on the momentum lattice as an order two shift in the momentum 
eigenvalue, $n$ $\to$ $n+\half$. Note that this shift vector is not thermal
duality invariant. We will adapt the left-right symmetric analysis of \cite{dgh} 
for the different ten-dimensional fermionic string theories and taking into
account the thermal boundary conditions. Thus, the action of 
the super-orbifold group will be required to preserve either one, 
or both, of the holomorphic N=1 superconformal invariances in, 
respectively, the heterotic, or type II, string theories, consistent 
with modular invariance and the correct thermal duality transformations. 

\vskip 0.1in
It is convenient to introduce lattice partition functions that transform 
straightforwardly under a thermal duality transformation \cite{dgh}. A
Poisson resummation makes their transformation properties under modular
transformations manifest. As in \cite{bosonic}, we introduce a 
dimensionless inverse temperature 
(radius) defining $x$ $\equiv$ $r(2/\alpha^{\prime})^{1/2}$, 
with $\beta$ $=$ $\pi (\alpha^{\prime}/2)^{1/2} x$. The 
dimensionless quantized momenta live in
a $(1,1)$ dimensional Lorentzian self-dual lattice 
\cite{nsw,dgh,polchinskibook}:
\begin{equation}
\Lambda^{(1,1)} :  \quad \quad\quad ({{\alpha^{\prime}}\over{2}})^{1/2}
(p_L , p_R) \equiv   (l_L , l_R) = ( 
{{n}\over{x}} + {{wx}\over{2}}   , 
{{n}\over{x}} - {{wx}\over{2}}  )  \quad , 
\label{eq:dimless}
\end{equation} 
with a natural decomposition into even and odd integer momentum 
(winding) sums. Thus, we define:
\begin{eqnarray} 
\Gamma^{++} (x) \equiv&& \sum_{w \in 2{\rm Z}, n \in 2{\rm Z}} 
 q^{{{1}\over{2}} ({{n}\over{x}} + {{wx}\over{2}} )^2 } 
 {\bar q}^{{{1}\over{2}} ({{n}\over{x}} - {{wx}\over{2}} )^2 } 
, \quad
\Gamma^{--} (x) \equiv \sum_{w \in 2{\rm Z}+1, n \in 2{\rm Z}+1} 
 q^{{{1}\over{2}} ({{n}\over{x}} + {{wx}\over{2}} )^2 } 
 {\bar q}^{{{1}\over{2}} ({{n}\over{x}} - {{wx}\over{2}} )^2 } 
 \cr
\nonumber
\Gamma^{+-} (x) \equiv&& \sum_{w \in 2{\rm Z}, n \in 2{\rm Z}+1} 
 q^{{{1}\over{2}} ({{n}\over{x}} + {{wx}\over{2}} )^2 } 
 {\bar q}^{{{1}\over{2}} ({{n}\over{x}} - {{wx}\over{2}} )^2 } 
, \quad
\Gamma^{-+} (x) \equiv \sum_{w \in 2{\rm Z}+1, n \in 2{\rm Z}} 
 q^{{{1}\over{2}} ({{n}\over{x}} + {{wx}\over{2}} )^2 } 
 {\bar q}^{{{1}\over{2}} ({{n}\over{x}} - {{wx}\over{2}} )^2 } 
 \quad . \cr
\quad && \quad
\label{eq:evenodd}
\end{eqnarray}
It is evident that the functions $\Gamma^{++}$ and $\Gamma^{--}$
are invariant while $\Gamma^{+-}$ is mapped to $\Gamma^{-+}$, and
vice versa, under both modular and thermal duality
transformations. 
For example, the following shift by half a lattice vector, 
$\delta$$=$$\half({{1}\over{x}} +{{x}\over{2}},{{1}\over{x}}-{{x}\over{2}})$,
giving shifted lattice partition functions:
\begin{equation} 
\Gamma_{\delta} (x) \equiv \sum_{w \in {\rm Z} + \half ,
n \in {\rm Z}+\half } 
 q^{{{1}\over{2}} ({{n }\over{x}} + {{wx}\over{2}} )^2 } 
 {\bar q}^{{{1}\over{2}} ({{n}\over{x}} - {{wx}\over{2}} )^2 } 
 \quad ,
\label{eq:evenoddshift}
\end{equation}
is invariant under a thermal duality transformation.
The shifted lattice can be further decomposed into even and odd 
lattices as above.
All of these considerations extend to lattices of higher dimension.
This generalizes the construction given in \cite{dgh}.

\vskip 0.6in
\section{Type IIA-IIB Strings at Finite Temperature}

We will now show that the vacuum functional for an ensemble of IIA or 
IIB strings takes identical form in the absence of gauge
fields, as a consequence of the thermal duality transformation mapping 
the thermal IIA string to the thermal IIB string. Although the 
thermal type II partition functions we derive will turn out to be 
pathological due to the appearance of tachyons--- over the entire finite 
temperature range as below, or above a string-scale minimum temperature
as in the next subsection, it is a useful warm-up exercise 
prior to the discussion of the tachyon-free heterotic string ensemble 
that follows. The behavior of the thermal type II ensemble at the 
critical temperatures when, respectively, the leading tachyonic, momentum
or winding, modes turn massless, might be of intrinsic interest for the 
understanding of the moduli space of string/M theory 
\cite{poltorus,aw,polchinskibook}\cite{aps}. 

\vskip 0.1in
The normalized generating functional of 
connected one-loop vacuum graphs for either type II 
string theory in the embedding space $R^9$$\times$$S^1/Z_2$ 
is given by a straightforward extension of \cite{bosonic}.
For either type II string, we have an expression of the form:
\begin{equation}
W_{\rm II} =  
\int_{\cal F} {{d^2 \tau}\over{4\tau_2^2}} 
  (2 \pi \tau_2 )^{-4} |\eta(\tau)|^{-14} Z_{\rm II} (\beta)
\quad ,
\label{eq:typeII}
\end{equation}
where the spatial volume, $V$ $=$ $ L^{9} (2\pi \alpha^{\prime})^{9/2} $,
and the inverse temperature is given by $\beta $ $=$ $ \pi r_{\rm circ.}$.

\vskip 0.1in
The function ${\rm Z}_{\rm II~ orb.}$ is the product of 
appropriate fermion
and boson partition functions: ${\rm Z}_F Z_{\rm B}$, to be
defined below.
As in the bosonic string, the $n$$=$$w$$=$$0$ sector is required to
be a subspace of the full thermal spectrum due to the necessity for
separate left and right-moving conserved charges on the world-sheet:
namely, the momentum modes labeled $+n$ and $-n$, with $n$ non-zero,
are both in the thermal spectrum,
and consequently, so is the $n$$=$$0$ mode. Likewise for winding states. 
The spectrum of thermal modes is unambiguously determined by modular 
invariance and by requiring that IIA winding modes are mapped
to IIB momentum modes, and vice versa, under a thermal duality
transformation. Thus, the vacuum functional of the IIA string, being
an intensive thermodynamic variable, is required to precisely map
into the vacuum functional of the IIB string under a thermal duality
transformation.

\vskip 0.1in
We will consider the action of the super-orbifold group $R (-1)^{N_F}$,
either with or an accompanying half-momentum or half-winding shift, 
$\delta$, in the one-dimensional Lorentzian momentum lattice. Modular
invariance requires that we include {\em both} momentum and winding 
modes in the type IIA, IIB, thermal partition functions. The phase in
the path integral is determined by the thermal duality transformation:
since IIA winding modes are mapped to IIB momentum modes, and vice
versa, under thermal duality, the phase factor in either type II string 
path integral is required to be thermal duality {\em invariant}. We emphasize
that this should not be misinterpreted as requiring self-duality of
the IIA or IIB ensemble; it follows as a consequence of requiring that
the vacuum functional for IIA maps to that for IIB.
In the presence of a Ramond-Ramond sector, the mapping between the
vacuum functionals is less straightforward. 

\vskip 0.1in
The purpose of the shift in the momentum lattice is to 
introduce a positive, temperature-dependent, 
shift in the spectrum of thermal masses. 
The spacetime supersymmetry breaking orbifold action $(-1)^{N_F}$ is
modified by the introduction of modular invariant, and temperature-dependent, 
phases. The phase factor must also be compatible with both the thermal 
duality transformations, as explained above, and also the requirement
that spacetime supersymmetry is restored in the zero temperature limit
of the vacuum functional, an idea originally proposed in \cite{aw}. 
Note that, for the thermal type II strings, due to the appearance of
a tachyonic state at a characteristic minimum temperature,
continuation of our final expression for the vacuum functional down to the 
supersymmetric zero temperature limit is formal, at best. This issue
will however be resolved upon introduction of gauge fields in the type II
strings via a non-trivial Ramond-Ramond sector. 

\vskip 0.1in
The action of R on $X^0$, and the computation for the bosonic R-orbifold
partition function, is reviewed in \cite{polchinskibook,bosonic}. The result 
is:
\begin{equation}
Z_{\rm orb.} = {{1}\over{2}} {{1}\over{\eta{\bar{\eta}}}} \left [ \sum_{w,n \in {\rm Z} } 
   q^{{{1}\over{2}} ({{n }\over{x}} + {{wx}\over{2}} )^2 } 
     {\bar q}^{{{1}\over{2}} ({{n}\over{x}} - {{wx}\over{2}} )^2 } 
  +  | \Theta_3 \Theta_4 | + | \Theta_2 \Theta_3 | + | \Theta_2 \Theta_4 |
 \right ] 
\quad .
\label{eq:bosod}
\end{equation}
The world-sheet fermions can be conveniently 
complexified into left- and right-moving
Weyl fermions. As in the superstring, the spin structures for all ten left- and 
right-moving fermions, $\psi^{\mu}$, ${\bar{\psi}}^{\mu}$, $\mu$ 
$=$ $0$, $\cdots$, $9$, are determined by those for the world-sheet gravitino 
associated with left- and right-moving N=1 superconformal invariances. 
Begin by recalling the spacetime supersymmetric
sum over spin structures familiar from the partition function of the
type II superstring \cite{polchinskibook}. We set 
${\rm Z}_F$$=$${\rm Z}_{SS}$, with:
\begin{equation}
 Z_{\rm SS} = 
     {{1}\over{4}}  {{1}\over{\eta^4{\bar{\eta}}^4}} 
  \left [ (\Theta_3^4 - \Theta_4^4 - \Theta_2^4)
   ({\bar{\Theta}}_3^4 - {\bar{\Theta}}_4^4 - {\bar{\Theta}}_2^4 )
\right ]
 \quad .
\label{eq:IIs}
\end{equation}
A factor of ${{1}\over{4}}$ arises from the average of spin structures 
in the type II theory \cite{polchinskibook}. Notice that the first of the relative signs 
in each round bracket preserves the tachyon-free condition. The second relative sign 
determines whether spacetime supersymmetry is preserved in the zero 
temperature spectrum. 
Under the action of the orbifold group $R \cdot (-1)^{N_F}$, 
ordinarily acting as $+1$ on spacetime 
bosons and as $(-1)$ on spacetime fermions, the  
phase of the contribution from spacetime fermions must be reversed 
as required by thermal boundary conditions. 
There is a unique choice of phases which implements this constraint
consistent with both modular invariance and broken supersymmetry at
arbitrary temperature:
we replace the theta functions with their absolute magnitudes, summing
over the different spin structures. 
This gives ${\rm Z}_F$$=$${\rm Z}_{NS}$, with:
\begin{equation}
 Z_{\rm NS} = 
     {{1}\over{4}}  {{1}\over{\eta^4{\bar{\eta}}^4}} 
  \left [ (|\Theta_3|^4 + |\Theta_4|^4 + |\Theta_2|^4)
   (|{\bar{\Theta}}_3|^4 + |{\bar{\Theta}}_4|^4 + |{\bar{\Theta}}_2|^4 )
 \right ] 
 \quad .
\label{eq:pfIItnsa}
\end{equation}
We now wish to generalize to an expression in which the spacetime 
supersymmetric and tachyon-free combination of spin structures 
given in Eq.\ (\ref{eq:IIs}) is recovered at zero temperature \cite{aw},
consistent with modular invariance and the thermal duality 
transformation interchanging IIA with IIB. 
Combining with the bosonic contributions, the required 
orbifold partition function describing {\em either}
the thermal IIA or IIB ensemble is a simple modification of that 
given in \cite{aw}:
\begin{eqnarray}
 Z_{\rm II } &&=
     {{1}\over{8}}  {{1}\over{|\eta|^{10}}} 
  \sum_{n,w\in {\rm Z}} \{ (|\Theta_3 |^8 
  + |\Theta_4|^8  + |\Theta_2|^8 ) \cr
  \quad&&\quad\quad + e^{ \pi i (n + w ) }
 [ (\Theta_2^4 {\bar{\Theta}}^4_4 + \Theta_4^4 {\bar{\Theta}}_2^4)
  - (\Theta_3^4 {\bar{\Theta}}_4^4 + \Theta_4^4 {\bar{\Theta}}_3^4 
   + \Theta_3^4 {\bar{\Theta}}_2^4 + \Theta_2^4 {\bar{\Theta}}_3^4 ) ] 
  \} q^{\half {\bf l}_L^2} {\bar {q}}^{\half {\bf l}_R^2} + \cdots 
 \quad , \cr
\quad && \quad
\label{eq:pfIItnsass}
\end{eqnarray}
where the ellipses denote the temperature independent states of the
R--orbifold.
This expression is modular invariant and transforms as desired 
under a thermal duality transformation: taking $\beta_{IIA}$$\to$$
$$\beta_{IIB}$$=$$\beta_C^2/\beta_{IIA}$, and simultaneously
interchanging the identification of winding and momentum modes,
$(n,w)_{IIA}$$\to$$(n'=w,w'=n)_{IIB}$.
The expression also has the desired zero 
temperature limits of the IIA and IIB strings: for 
large $\beta_{IIA}$, terms with $w_{IIA}$$\neq$$0$
decouples in the double summation since they are exponentially 
damped. The remaining terms are resummed by a Poisson 
resummation, thereby inverting the $\beta_{IIA}$ dependence in the 
exponent, and absorbing the phase factor in a shift of argument.
The large $\beta_{IIA}$ limit can then be taken smoothly 
providing the usual momentum integration for a non-compact dimension,
plus a factor of $\tau_2^{-1/2}$ from the Poisson resummation.
This recovers the supersymmetric zero temperature partition function.
A thermal duality transformation maps small $\beta_{IIA}$ to large 
$\beta_{IIB}$$=$$\beta_C^2/\beta_{IIA}$, also interchanging the
identification of momentum 
and winding modes. We can analyze the limit of large dual
inverse temperature as before, obtaining the zero temperature 
limit of the dual IIB theory.
At any intermediate temperature, all of the
thermal modes contribute to the vacuum functional with
a phase that takes values $(\pm 1)$ only. Note that the spacetime 
fermions of the zero temperature spectrum now contribute with
a reversed phase, evident in the first term in 
Eq.\ (\ref{eq:pfIItnsass}), as required by the
thermal boundary conditions.
The non-trivial winding and momentum modes in the thermal 
spectrum have no zero temperature counterpart; they are 
necessitated by modular invariance of the string spectrum. 

\vskip 0.1in
Note that while this expression has the expected benign spacetime 
supersymmetric zero temperature limits,
the physical state spectrum is replete with thermal tachyons 
over the entire finite temperature range 
as in the closed bosonic string \cite{bosonic}. It 
describes pathological, and physically unacceptable, thermal 
type II string theories. The pathological behavior over the 
entire {\em finite} temperature range of the expression in 
Eq.\ (\ref{eq:pfIItnsass}) is also true of the expression for 
the vacuum functional of the thermal type II string proposed 
in Eq.\ (5.20) of reference \cite{aw}, 
An important difference is that due to the momentum and 
winding number dependent phases introduced in the sum 
over spin structures, the expression for the type II vacuum 
functional given in \cite{aw} does {\em not} transform 
correctly under a thermal duality transformation.
Namely, while spacetime supersymmetry 
holds in the zero temperature limit, a thermal duality 
transformation mapping 
$\beta_{IIA}$$\leftrightarrow$$\beta_{IIB}$$=$$\beta_C^2/\beta_{IIA}$
gives a function that is tachyonic and nonsupersymmetric in
the (dual) zero temperature limit. This behavior was explicit in
Eq.\ (5.20) of \cite{aw}. 

\vskip 0.1in
Although we have presented this discussion for pedagogical purposes,
we emphasize that the tachyonic type II partition functions discussed 
in this subsection describe inherently {\em unstable} thermal ground 
states. We will dismiss them as pathological, and
modify our type II orbifold construction as described next.

\subsection{Minimum Temperature for the Thermal Type II Strings}

\vskip 0.1in
A non-pathological thermal type II vacuum functional
describing a {\em tachyon-free spectrum within a finite, 
albeit string-scale, temperature range}, and transforming
correctly under a thermal duality transformation, is
obtained as follows. Consider the duality invariant (1,1) lattice
partition function obtained by shifting each (n,w) vector 
by the constant vector, $\delta $, composed from a half-winding 
plus a half-momentum shift. The following expression is modular
invariant. It also transforms correctly under thermal duality. 
We replace the function $Z_{\rm II}$ appearing in 
Eqs.\ (\ref{eq:typeII}), (\ref{eq:pfIItnsass}) with:
\begin{eqnarray}
 Z_{\rm II} &&=
     {{1}\over{4}}  {{1}\over{|\eta|^{10}}} 
  \sum_{n,w\in {\rm Z}+\half } [ (|\Theta_3 |^8 
  + |\Theta_4|^8  + |\Theta_2|^8 ) \cr
  \quad&&\quad\quad + e^{ \pi i (n + w ) }
 \{ (\Theta_2^4 {\bar{\Theta}}^4_4 + \Theta_4^4 {\bar{\Theta}}_2^4)
  - (\Theta_3^4 {\bar{\Theta}}_4^4 + \Theta_4^4 {\bar{\Theta}}_3^4 
   + \Theta_3 {\bar{\Theta}}_2^4 + \Theta_2 {\bar{\Theta}}_3^4 ) \}
  ] q^{\half {\bf l}_L^2} {\bar {q}}^{\half {\bf l}_R^2} 
+ \cdots 
\cr
\quad && 
\label{eq:pfIItnsassd}
\end{eqnarray}
describing either type IIA or type IIB thermal strings.
The IIA and IIB descriptions are interchangeable, related by a 
thermal duality transformation:
$\beta_{IIA}$$\leftrightarrow$$\beta_{IIB}$$=$$\beta_C^2/\beta_{IIB}$,
simultaneously interchanging momentum and winding modes.
This partition function describes non-supersymmetric theories, 
except in the zero temperature limit of large radius.

\vskip 0.1in
The expression in Eq.\ (\ref{eq:pfIItnsassd}) describes a viable 
candidate world-sheet superconformal field theory satisfying the 
required consistency conditions for a weakly coupled ground 
state of the type IIA, or IIB, string theories. There is an obvious
underlying worldsheet representation of the lattice partition
function by a pair of chiral bosons. We will now show 
that Eq.\ (\ref{eq:pfIItnsassd}) gives a physically meaningful 
vacuum functional describing
the equilibrium thermal behavior of a stable ensemble of type II strings. 
To check for potential tachyonic instabilities, consider the mass 
formula in the (NS,NS) sector for world-sheet fermions,
with ${\bf l}_L^2 $ $=$ $ {\bf l}_R^2$, 
and $N$$=$${\bar{N}}$$=$$0$:
\begin{equation}
({\rm mass})^2_{nw} =  {{2}\over{\alpha^{\prime}}} 
 \left [ - 1 +  
{{\alpha^{\prime} \pi^2(n+\half)^2 }\over{2 \beta^2}} 
+ {{\beta^2(w+\half)^2}\over{2 \pi^2 \alpha^{\prime} }}  \right ]  
\quad .
\label{eq:massII}
\end{equation}
This is the only sector that can contribute tachyons to the thermal
spectrum. A nice check is that the momentum dependent phase factor 
introduced above does not enter the NS-NS sector; all potentially 
tachyonic states are spacetime scalars as expected.
Notice that the $n$$=$$w$$=$$0$ sector common to both type II
string theories contains a potentially tachyonic state whose 
mass is now temperature dependent. However, it corresponds to 
an unphysical tachyon. The potentially tachyonic
physical states are the pure momentum and pure winding states, 
$(n,0)$ and $(0,w)$, with $N$$=$${\bar{N}}$$=$$0$. The R-orbifold
projects to the symmetric linear combinations of net zero
momentum and net zero winding number states as explained in 
\cite{bosonic}.

\vskip 0.1in
The effect of the shift in the momentum lattice is to open up
a window on the temperature axis, for which the physical state 
spectrum is tachyon-free. Analogous to the bosonic string analysis, 
we can compute the temperatures
below, and beyond, which these modes become tachyonic in the absence
of oscillator excitations. Each momentum mode, 
$(\pm n, 0)$, is tachyonic {\em upto} some critical temperature, 
$T_n^<$ $=$ $(2/\alpha^{\prime})^{1/2}/\pi (n+\half) $, after which it 
becomes stable. Conversely, each winding mode $(0,\pm w)$, is 
tachyonic {\em beyond} some critical temperature, 
$T_w^>$ $=$ $(w+\half )/\pi (2\alpha^{\prime})^{1/2}$.
It is evident that the
$(1,0)$ and $(0,1)$ states determine the upper and lower 
critical temperatures for a tachyon-free physical state 
spectrum. We refer to these, respectively, as the minimum and maximum
temperatures for the given type II string.
A thermal duality transformation maps IIA to IIB, 
mapping $T_{\rm min}(IIA)$ to $T_{\rm max}(IIB)$, 
and vice versa,
since the identification of momentum and winding modes is
switched in the mapping.
Expressed in terms of the dual temperature
variable, the minimum and maximum temperatures will, of 
course, coincide. As a consequence, the IIA
and IIB thermal strings have common minimum and 
maximum temperatures.
The thermal mass spectrum of either type II string is 
tachyon-free for temperatures in the range: 
\begin{equation}
T_{\rm min} < T < T_{\rm max} , \quad {\rm where} 
 \quad T_{\rm min} = 2{\sqrt{2}}/3\pi\alpha^{\prime 1/2},
\quad T_{\rm max} = 3/2{\sqrt{2}}\pi\alpha^{\prime 1/2} 
 \quad . 
\label{eq:rant}
\end{equation}
Notice that there are no additional massless thermal modes at the
self-dual temperature, $\beta_C^2$$=$$\pi^2 \alpha'$, from which
we can infer
that only one fixed line of the world-sheet superconformal
field theory passes through the self-dual point, parameterized by the
inverse temperature. The thermal type II strings are stable at
least in the vicinity of the self-dual temperature. 

\vskip 0.1in
The Kosterlitz-Thouless transition at $T_C$ is a continuous phase
transition mapping the thermal type IIA string to the thermal IIB string. 
On the other hand, at the minimum and maximum temperatures, respectively, 
the $(1,0)$ and $(0,1)$ modes which are massive within the allowed 
temperature regime, turn tachyonic. They represent instabilities 
of the thermal type II string phase.\footnote{It is tempting to
consider the possibility of a phase transition at which a closed 
string tachyon condensate drives the tachonic type II vacuum to a new 
tachyon-free ground state, possibly even the supersymmetric
zero temperature ground state, as conjectured for non-compact ALE spaces
in the leading order in $\alpha^{\prime}$ analysis of \cite{aps}. 
Unfortunately, our result for $T_{\rm min}$ shows that this scenario 
would appear to violate Zamolodchikov's c-theorem in a regime of 
strict validity \cite{zam}: the conjectured 
transition would have to occur at {\em finite}, string-scale, temperature, 
in the {\em compact} target space, 
$R^9$$\times$$S^1/Z_2$, of finite volume, 
$\pi \alpha^{\prime 1/2}$. A tachyonic vertex
operator corresponds to a relevant perturbation of the world-sheet 
superconformal field theory and the RG flow is towards a 
fixed point of lower central charge.
In the manifestly Weyl-invariant framework of critical string theory this 
is forbidden without nontrivial dynamics for the super-Liouville mode 
\cite{polyakov,ddk,dk,jp}. 
A possible subtlety is that the putative closed string 
tachyon condensate is unlikely to correspond to a local operator in the 
world-sheet superconformal field theory. In this event, the c-theorem would
not apply and cannot be used to rule out such a transition. In any case,
a simple resolution for the thermal type II string is the inclusion of a 
Ramond-Ramond sector: thermal tachyonic modes are eliminated at all 
temperatures starting from zero upon turning on a temperature dependent 
timelike Wilson line, 
and the moduli space will be free of tachyonic ground states.}
Consequently, the tachyon-free thermal type II strings described here 
are {\em inherently finite temperature} nonsupersymmetric ground 
states of string/M theory. The thermal ground state is physically
sensible only for a background temperature of order the 
string scale, $T$$\simeq$$T_C$$=$$1/\pi\alpha^{\prime 1/2}$. 
This is unambiguous indication that flat supersymmetric 
10D spacetime without gauge fields is an unstable perturbative 
ground state for both type IIA and type IIB string theories \cite{us}. 
In the conclusion, we point out that a rather simple resolution of
this sickness of type II string theory does exist, but it requires the 
inclusion of potential gauge fields from a non-trivial Ramond-Ramond 
sector. 

\vskip 0.1in
It is interesting that the source of this instability is
distinct from that found in the finite temperature field 
theoretic analysis of \cite{gyp}. The problem addressed 
here is that of the internal consistency required of a 
non-pathological, and {\em modular invariant}, thermal free string 
spectrum: the thermal strings do not gravitate at this 
order in string perturbation theory, and the origin of our 
no-go result should not be confused with the gravitational 
instability of flat space at finite temperature.

\section{The 10D Heterotic Superstrings at Finite Temperature}
 
Unlike the type II string which has no Yang-Mills fields in the
absence of a Ramond-Ramond sector and, consequently, no weak 
coupling tachyon-free nonsupersymmetric ground states at low 
temperatures, flat and tachyon-free nonsupersymmetric ground states 
are known to exist in the perturbative heterotic string theory 
\cite{sw,agmv,dh,klt,dienes}. The dilaton one-point function 
is likely to
be non-vanishing in such a vacuum, and it will be important to 
verify that the constant dilaton one-point function can consistently 
be absorbed in a renormalization of the bare string tension 
\cite{wein,rg}. This is discussed further in the conclusions.
In the event of a non-constant one-point function for dilaton, 
or graviton, an inescapable flow to strong coupling would be
inevitable.  However, since our interest is in the equilibrium 
behavior of free heterotic strings, we will ignore such loop 
effects for the moment and consider the possibilities for a 
tachyon-free finite temperature ground state in the heterotic 
string theory.

\vskip 0.1in
We will find that
the 10D $E_8$$\times$$E_8$ and $SO(32)$ heterotic superstrings 
share a stable and tachyon-free
ground state at all temperatures starting from zero in the 
presence of a temperature-dependent Wilson line background,
with non-abelian gauge group $SO(16)$$\times$$SO(16)$. The crucial
difference from type II string theory is the availability of gauge 
fields at weak coupling \cite{nsw,ginine,polchinskibook}.
It is interesting that even an infinitesimal change in the 
background temperature results in some of the massless vector bosons
of the 10D gauge theory acquiring masses of order the supersymmetry
breaking scale which, in this case, is the string scale, 
$\alpha^{\prime -1/2}$. This unexpected 
result is a consequence of the temperature-dependent Wilson line, 
necessitated by requiring a stable, and tachyon-free, thermal
ensemble of heterotic strings.

\vskip 0.1in
We consider an equilibrium ensemble of free heterotic 
strings in nine noncompact
dimensions occupying the box-regulated spatial volume $V$ $=$ 
$L^9 (2\pi \alpha^{\prime})^{9/2}$. The normalized
vacuum functional, or the
generating functional of connected one-loop
vacuum string graphs, is given by:
\begin{equation}
W_{\rm het.} = 
\int_{\cal F} {{d^2 \tau}\over{4\tau_2^2}} 
  (2 \pi \tau_2 )^{-9/2} |\eta(\tau)|^{-14} Z_{\rm het.} (\beta)
\quad ,
\label{eq:het}
\end{equation}
where the precise action of the super-orbifold group on both world-sheet, 
and gauge, fermions that determines the form of $Z_{\rm het.}$ remains to be 
specified. The free energy of the free heterotic string gas
can be obtained from the usual relation: $F$$=$$-W_{\rm het.}/\beta$.
We begin with the modular invariant and 
spacetime supersymmetric vacuum 
functional of the ten-dimensional $E_8$$\times$$E_8$ heterotic string: 
\begin{equation}
W_{\rm het.}|_{\beta = \infty} = 
\int_{\cal F} {{d^2 \tau}\over{4\tau_2^2}} 
  (2 \pi \tau_2 )^{-5} |\eta(\tau)|^{-16}
{{1}\over{8}}  {{1}\over{{\bar{\eta}}^4}} 
   ({\bar{\Theta}}_3^4 - {\bar{\Theta}}_4^4 - {\bar{\Theta}}_2^4 )
\left [ ({{\Theta_3}\over{\eta}})^8 
+ ({{\Theta_4}\over{\eta}})^8 
+ ({{\Theta_2}\over{\eta}})^8 \right ]^2 
\quad .
\label{eq:holos}
\end{equation}
The partition function in Eq.\ (\ref{eq:holos})
is of the form $Z_{\rm S} (Z_8)^2$, where
$Z_S$ is the holomorphic partition function given by the spacetime
supersymmetric sum over chiral spin structures, and $(Z_8)^2$ is
the contribution of the $32$ gauge fermions. In bosonic form, $Z_8^2$
may be expressed as the lattice partition function for the 
16D $E_8$$\oplus$$E_8$ Euclidean self-dual lattice 
\cite{polchinskibook}.
As in the thermal type II case, we wish to identify a
a nonsupersymmetric and tachyon-free partition function 
at finite temperature, $Z_{\rm het}(\beta)$,
appropriate for substitution in Eq.\ (\ref{eq:het}). 
$Z_{\rm het.}$ describes the mass spectrum of free 
$E_8$$\times$$E_8$ strings at finite temperature. It is 
required to be both modular invariant, and also
self-dual under thermal duality transformations, 
reverting to the supersymmetric partition function,
$Z_S (Z_{\rm 8})^2$, in the limit of zero temperature.
Finally, as evident from the analysis in \cite{ginine}, 
the vacuum functional of the 
finite temperature $E_8$$\times$$E_8$ heterotic string is
expected to map precisely into the vacuum functional
of the finite temperature $SO(32)$ heterotic string
under appropriate $SO(17,1)$ transformations.

\vskip 0.1in
We will consider the action of the super-orbifold group 
$R(-1)^{N_F}$ accompanied by a temperature dependent
Wilson line in the $E_8$$\oplus$$E_8$ lattice. 
The reflection, $R$, has been considered at length
in both the bosonic case \cite{bosonic}, and in earlier sections. 
Let us focus on understanding the Wilson line background. Since 
$(-1)^{N_F}$ acts trivially on the timelike coordinate, $X^0$, and the 
$16$ current algebra bosons, the result of a temperature-dependent 
timelike Wilson line follows from the analysis of supersymmetric
9D heterotic strings in \cite{ginine}. Namely, we compactify the 10D 
$E_8$$\times$$E_8$ supersymmetric string
on a circle of radius $r_{\rm circ.}$. 
Introduction of a radius-dependent Wilson line background, 
${\bf A}$$=$${{2}\over{x}}(1,0^7,-1,0^7)$, $x$$=$$
({{2}\over{\alpha^{\prime}}})^{1/2} r_{\rm circ.}$,
interpolates continuously between a 9D supersymmetric
$SO(16)$$\times$$SO(16)$ string at generic radii and the 
non-compact limit where the gauge group is enhanced to 
$E_8$$\times$$E_8$. Note that the states in the spinor 
lattices of $SO(16)$$\times$$SO(16)$ provide additional 
massless vector bosons only in the limit of infinite 
$x$. Consider the $(17,1)$-dimensional
lattice, $E_8$$\oplus$$E_8$$\oplus$$U$, where $U$ is the 
$(1,1)$ lattice that describes compactification 
on a circle of radius 
$r_{\rm circ.}$$=$$x(\alpha^{\prime}/2)^{1/2}$. A generic Wilson
line corresponds to the lattice boost \cite{ginine}:
\begin{equation}
({\bf p}; l_L, l_R) \to 
({\bf p}'; l^{\prime}_L, l^{\prime}_R) = 
({\bf p} + w x {\bf A};
u_L - {\bf p}\cdot {\bf A}
- {{wx}\over{2}} {\bf A} \cdot {\bf A} ,
u_R - {\bf p}\cdot {\bf A}
- {{wx}\over{2}} {\bf A} \cdot {\bf A} ) \quad .
\label{eq:boost}
\end{equation}
${\bf p}$ is a 16-dimensional lattice vector in 
$E_8$$\oplus$$E_8$. The vacuum functional of the 9D 
supersymmetric theory, with generic radius and
generic Wilson line in the compact spatial direction,
can accordingly be written in terms of a sum over
vectors in the boosted lattice:
\begin{equation}
W_{\rm het.} (r_{\rm circ.} ; {\bf A}) = 
\int_{\cal F} {{d^2 \tau}\over{4\tau_2^2}} 
  (2 \pi \tau_2 )^{-5} |\eta(\tau)|^{-16}
{{1}\over{8}}  {{1}\over{{\bar{\eta}}^4}} 
   ({\bar{\Theta}}_3^4 - {\bar{\Theta}}_4^4 - {\bar{\Theta}}_2^4 )
\left [ {{1}\over{\eta^{16}}}
\sum_{({\bf p}'; l^{\prime}_L , l^{\prime}_R)} q^{\half ({\bf p}^{\prime 2} +
  l_L^{\prime 2})} {\bar q}^{\half l_R^{\prime 2}} 
\right ]
\quad .
\label{eq:hollat}
\end{equation}

\vskip 0.1in
The corresponding compactification of the 10D $E_8$$\times$$E_8$
string on the spacetime $R^9$$\times$$S^1/Z_2$
with Euclidean signature, 
describes the equilibrium thermal spectrum of $E_8$$\times$$E_8$
strings at finite temperature. The interval length of Euclidean
time is the inverse temperature, with $x$$=$$
({{2}\over{\alpha^{\prime}}})^{1/2} {{\beta}\over{\pi}}$.
Turn on the temperature dependent Wilson line:
${\bf A}$$=$${{2}\over{x}}{\rm diag}(1,0^7,-1,0^7)$
\cite{ginine}, where ${\bf A}$ is the timelike component
of the vector potential, accompanied  
by $R(-1)^{N_F}$ acting on the ten right-moving world-sheet 
fermions. The result will be a nonsupersymmetric, but tachyon-free, 
9d $SO(16)$$\times$$SO(16)$ heterotic string. 
We should clarify that, although it is not apparent in the
form of the Wilson line, the Lorentzian self-dual boosted lattice 
is in fact invariant under a thermal duality transformation.

\vskip 0.1in
Notice that, from the viewpoint of the low-energy
finite temperature gauge theory, the timelike Wilson line
can be understood as imposing a modified axial gauge
condition: $A^0$$=$${\rm const}$.
The dependence of the constant on background temperature is 
chosen so as to provide a shift in the string mass spectrum 
that precisely compensates for potential low temperature 
$(n,0)$ tachyonic modes. Such a special gauge choice may
surprise the reader. We note that from the viewpoint of
dynamics in the finite temperature gauge theory, the value
of this constant is of no consequence. 
On the other hand, from the viewpoint of 
heterotic string theory, 
${\bf A}$$=$${{2}\over{x}}{\rm diag}(1,0^7,-1,0^7)$,
$0$$\leq$$x$$\leq$$\infty$, describes a one-parameter 
direction in field space along which supersymmetry is 
spontaneously broken at the string scale. While possibly not
of phenomenological interest, it is remarkable that the 
mass spectrum of the resulting line of nonsupersymmetric string 
theories is {\em tachyon-free at all temperatures starting 
from zero}. This is clear evidence that 10D N=1 supersymmetric flat 
space with either $E_8$$\times$$E_8$ or $SO(32)$ Yang-Mills 
gauge fields is a tachyon-free heterotic string
ground state at weak coupling, 
under both infinitesimal, and finite, variation in the 
background temperature. At finite temperature, we have a 
spontaneous breaking of both supersymmetry and the
non-abelian gauge symmetry. It would be of great interest to 
understand the dynamics underlying this phenomenon.

\vskip 0.1in
It remains to identify the form of the interpolating function
$Z_{\rm het.}(\beta)$ that describes the modular invariant partition 
function of the $SO(16)$$\times$$SO(16)$ theory,
such that the $E_8$$\times$$E_8$ partition function given in eq.\
(\ref{eq:holos}) is recovered in the zero temperature limit.
In addition, the vacuum functional is required to be self-dual
under thermal duality transformations, with simultaneous 
interchange of thermal momentum and thermal winding modes.
Consider the modular invariant function:
\begin{eqnarray}
Z_{\rm het.} =&& {{1}\over{4}}
\sum_{n,w} 
\left [ ({{\Theta_2}\over{\eta}})^8
({{\Theta_4}\over{\eta}})^8 
({{{\bar{\Theta_3}}}\over{\eta}})^4 
- ({{\Theta_2}\over{\eta}})^8
({{\Theta_3}\over{\eta}})^8
({{{\bar{\Theta_4}}}\over{\eta}})^4 
- ({{\Theta_3}\over{\eta}})^8
({{\Theta_4}\over{\eta}})^8 
({{{\bar{\Theta_2}}}\over{\eta}})^4 \right ] 
q^{\half {\bf u}^2_L}
{\bar{q}}^{\half {\bf u}_R^2} \cr
&& \quad + 
{{1}\over{2}} \sum_{n,w} 
e^{\pi i (n+w)} \left [ ({{{\bar{\Theta_3}}}\over{\eta}})^4 
({{\Theta_3}\over{\eta}})^8 \{ ({{\Theta_2}\over{\eta}})^8 +
({{\Theta_4}\over{\eta}})^8 \} + \cdots \right ] 
q^{\half {\bf u}_L^2}
    {\bar{q}}^{\half {\bf u}_R^2} + {\rm other} 
\quad .
\cr
\quad && \quad
\label{eq:ident}
\end{eqnarray}
The ellipses denote symmetrization among the three Jacobi theta functions. 
\lq\lq Other" denotes the temperature independent contributions from the
fixed points of the $R$-orbifold discussed in \cite{bosonic}. 
Note that taking the $x$$\to$$\infty$ 
limit, by similar manipulations as in the type II case, yields the 
partition function of the supersymmetric 10D $E_8$$\times$$E_8$ string. 
The Wilson line becomes irrelevant in the non-compact limit. 
We can follow the $SO(17,1)$ transformation described above with a 
lattice boost decreasing the size of the interval \cite{ginine}:
\begin{equation}
 e^{-\alpha_{00}} = {{1}\over{1+|{\bf A}|^2/4}} \quad .
\label{eq:scale}
\end{equation}
This recovers the Spin(32)/Z$_2$ theory compactified on an interval 
of size $2/x$, but with Wilson line ${\bf A}$$=$$x{\rm diag}(1^8,0^8)$ 
\cite{ginine}. Thus, taking the large radius limit in the dual variable, 
with dual Wilson line background, gives instead the 10D $Spin(32)/Z_2$ 
supersymmetric heterotic string. The thermal $E_8$$\times$$E_8$ 
and $SO(32)$ heterotic string are found to share the same 
tachyon-free finite temperature ground state. The Kosterlitz-Thouless
transformation at $T_C$$=$$1/\pi\alpha^{\prime 1/2}$ is a self-dual 
continuous phase transition in this theory.
We comment that combining the $R(-1)^{N_F}$ action with modding
out by the outer automorphism exchanging the two $E_8$ lattices 
gives a distinct tachyon-free ground state with gauge group
$E_8$ which is discussed elsewhere \cite{gv,chl,us,type1}.

\vskip 0.1in
It is interesting to compare with a different noncompact limit of the
nonsupersymmetric and tachyon-free 9d $SO(16)$$\times$$SO(16)$ 
theory: consider directly taking $\beta$$\to$$\infty$ with no corresponding 
action on the world-sheet fermions. Notice that although the Wilson line 
above becomes trivial in the noncompact limit, the approach to 10D occurs
by a distinct sequence of continuously connected $SO(17,1)$ transformations: 
in this case, simple radius change. As a consequence, the 10D theory is 
{\em nonsupersymmetric}, tachyon-free, and with unchanged gauge group,
defined on a Euclidean spacetime $R^{9}$$\times$$S^1/Z_2$, where the
$Z_2$ denotes the bosonic R-orbifold.
This theory describes the finite temperature behavior
of equilibrium nonsupersymmetric $SO(16)$$\times$$SO(16)$ heterotic 
strings living in a 10D flat spacetime with Lorentzian signature. 
This Lorentzian solution was first discovered in \cite{agmv}, and 
gives the 
unique modular invariant, nonsupersymmetric, and tachyon-free, 10D 
heterotic string partition function. The thermal partition function 
of the 9D theory takes the simple form: 
\begin{equation}
Z_{\rm NS} = {{1}\over{4}}
\left [ ({{\Theta_2}\over{\eta}})^8
({{\Theta_4}\over{\eta}})^8 
({{{\bar{\Theta_3}}}\over{\eta}})^4 
- ({{\Theta_2}\over{\eta}})^8
({{\Theta_3}\over{\eta}})^8
({{{\bar{\Theta_4}}}\over{\eta}})^4 
- ({{\Theta_3}\over{\eta}})^8
({{\Theta_4}\over{\eta}})^8 
({{{\bar{\Theta_2}}}\over{\eta}})^4 \right ]
  \sum_{n,w} q^{\half {\bf l}_L^2 } {\bar{q}}^{\half {\bf l}_R^2} 
+ \cdots
\quad ,
\label{eq:dualh}
\end{equation}
where the ellipses denote temperature independent contributions from
the fixed points of the R-orbifold. The term in square brackets is the
partition function of the 10D nonsupersymmetric and tachyon-free string 
found in \cite{agmv}. The $(1,1)$-d lattice partition function 
containing thermal momentum and winding modes is completely decoupled.
The thermal self-duality transition at the Kosterlitz-Thouless point
in this theory is identical to that in the closed bosonic string.
The only distinction is that the string vacuum functional, the
Helmholtz and Gibbs free energies, the internal energy,
and all subsequent thermodynamic potentials, are both finite and 
tachyon-free. Notice that the low energy finite temperature gauge 
theory is now quantized in ordinary axial gauge, ${\bf A}^0$$=$$0$: 
the finite temperature limit preserves both the nonabelian gauge 
symmetry and the (non)supersymmetry of the zero temperature limit. 

\vskip 0.1in
Consider instead an orbifold compactification of the 10D nonsupersymmetric 
$SO(16)$$\times$$SO(16)$ string, acting with 
the order two twist, $e^{2\pi i {\bf \delta} \cdot {\bf p}}$,
where $\delta$ is a half-winding lattice vector. This was referred 
to as the twist I orbifold in \cite{itoyama}. It
was shown to have a small radius limit in which spacetime 
supersymmetry is asymptotically restored with 
simultaneous enhancement of the 
gauge group to $E_8$$\times$$E_8$. This is the {\em reverse} of 
the phenomenon we considered above. It may be interpreted as the 
spontaneous restoration of symmetries at high temperatures, 
with asymptotic approach to zero radius. We should clarify that
the result of a thermal duality transformation on the vacuum 
functional of the self-dual twist I orbifold, 
will give the vacuum functional 
of the dual twist II orbifold of \cite{itoyama},
where twist II refers to $\delta$ a half-momentum lattice vector.
Written in the dual variable,
the vacuum functional demonstrates a spontaneous restoration of 
supersymmetry with enhancement of gauge group to $E_8$$\times$$E_8$
in the large dual radius, or zero temperature, limit. 
Presumably, the 9D nonsupersymmetric and tachyon-free 
$SO(16)$$\times$$SO(16)$ string can also be obtained as an orbifold 
compactification of the 10D $SO(32)$ heterotic superstring. This 
would complete the correspondence with the circle of dualities 
described above. What underlies this correspondence is the well-known 
equivalence of quantized Wilson line backgrounds with a shift in the 
periodicity, or world-sheet boundary condition, of the chiral bosons 
in the current algebra underlying the gauge lattice of the heterotic 
string. The description in terms of Wilson line backgrounds \cite{ginine}
gives a clearer picture of the connectivity of the moduli space of the
finite temperature theory.

\vskip 0.1in
The Helmholtz free energy of the tachyon-free gas of free heterotic
strings follows from the definition below Eq.\ (\ref{eq:het}), and
is clearly finite at $T_H$, with no evidence for a Hagedorn divergence
and, consequently, no Hagedorn phase transition.   
The internal energy of the ensemble of free $SO(16)$$\times$$SO(16)$ 
heterotic strings can be computed as follows. Recalling the thermodynamic 
relations and definitions given in \cite{bosonic}, we
obtain the following expression for the internal energy:
\begin{equation}
 U = - \left ( {{\partial W}\over{\partial \beta }} \right )_V =
 \half \int_{\cal F}
{{|d\tau|^2}\over{4\tau_2^2}} (2\pi\tau_2)^{-9/2}
   |\eta(\tau)|^{-16} 
{{4\pi \tau_2}\over{\beta}}
\sum_{n,w \in {\rm Z} }
  \left ( {{w^2 x^2}\over{4}} - {{n^2}\over{x^2}}  \right )
\cdot q^{ {{1}\over{2}}{\bf l}_L^2 }
     {\bar q}^{{{1}\over{2}}{\bf l}_R^2 }
\cdot Z_{\rm [SO(16)]^2} \quad ,
\label{eq:term}
\end{equation}
where $Z_{\rm [SO(16)]^2}$ denotes the factor in square brackets
appearing in, respectively, Eq.\ (\ref{eq:ident}), or Eq.\ (\ref{eq:dualh}).
The two results differ in the noncompact zero temperature limit 
giving, respectively, the supersymmetric $E_8$$\times$$E_8$,
or nonsupersymmetric $SO(16)$$\times$$SO(16)$, 10D string theories.
We emphasize that {\em an unambiguous specification of the zero 
temperature limit of the finite temperature heterotic string 
requires specifying a possible Wilson line background in the 
timelike direction}. From the viewpoint of the low energy 
gauge-gravity theory, this is the assertion that recovering the 
supersymmetric zero temperature limit requires imposing a 
particular axial 
gauge choice for the finite temperature gauge-gravity theory.

\vskip 0.1in
The thermodynamic potentials can be computed using the same 
methods as shown 
for the self-dual--- but tachyonic, ensemble of free 
closed bosonic strings in \cite{bosonic}. It is evident that 
taking
successive partial derivatives of the string effective action 
functional yields an {\em infinite} hierarchy of analytic 
functions of the inverse temperature. 
Each of the thermodynamic potentials displays a continuous 
phase transition at the self-dual temperature, unambiguously
identifying a phase transition of the Kosterlitz-Thouless
type \cite{kosterlitz,bosonic}. Of course, unlike the ensemble of 
closed bosonic strings, the thermodynamic potentials of the 
heterotic ensemble are {\em finite} normalizable functions 
at all temperatures starting from zero. 
We emphasize that, unlike the self-dual
vacuum functional, the expressions for the Helmholtz free energy 
and generic thermodynamic potentials are {\em not} invariant 
under thermal duality transformations. 
The vacuum functional of the $SO(16)$$\times$$SO(16)$ string 
is known to be {\em negative} \cite{agmv},
due to the preponderance of spacetime fermionic modes over
spacetime bosonic modes. The internal energy of free 
heterotic strings is negative 
at low temperatures, vanishing precisely at the self-dual 
temperature $x_c$$=$$2^{1/2}$,
$T_c$$=$$1/\pi \alpha^{\prime 1/2}$. 
The Helmholtz, or Gibbs, free energies are minimized at
the self-dual temperature. As is physically reasonable
for a stable thermodynamic ensemble at equilibrium \cite{aw}, 
the internal energy is found to be
a monotonically increasing function of temperature.

\section{Conclusions}

We have shown that the clarification of the connectivity of the moduli 
space of nine dimensional heterotic string ground states given in 
\cite{nsw,ginine} enables interpretation of the tachyon-free, and 
nonsupersymmetric, 9D $SO(16)$$\times$$SO(16)$ heterotic theory as the 
finite temperature ground state shared by the 10D $E_8$$\times$$E_8$
and $SO(32)$ heterotic superstrings \cite{us}. 
The crucial difference in the type II strings 
is the fact that, in the absence of gauge fields, 
even an infinitesimal variation in background temperature away 
from the zero temperature vacua--- the 10D supersymmetric 
flat spacetime IIA or IIB strings,
results in low temperature tachyonic modes in the string spectrum.
For the 10D type I and heterotic string theories, on the other hand,
the presence of gauge fields enables turning on a temperature dependent 
timelike Wilson line, resulting in a tachyon-free and nonsupersymmetric 
string spectrum at all temperatures starting from zero. The tachyon-free
finite temperature string theories have nonabelian gauge group 
$SO(16)$$\times$$SO(16)$ \cite{us,type1}. 

\vskip 0.1in
We should emphasize that the absence of tachyonic modes 
in the string mass spectrum resulting from an
infinitesimal change in the background temperature away 
from zero, is a crucial {\em
low temperature} consistency condition for any phenomenologically viable 
supersymmetric String/M theory ground state describing our Universe. 
Surprisingly, this
concern has not received much attention in the String/M phenomenology 
literature. Finite temperature necessarily breaks supersymmetry. Not 
surprisingly, one or other massless modulus, including the dilaton field, 
could develop a potential with subsequent evolution to 
imaginary eigenvalues in the mass matrix. But our analysis of free, 
noninteracting, closed strings shows that the tachyonic thermal 
instabilities have a different origin--- with no obvious 
correspondence in the low energy effective field theory.
They are a consequence of a well-established, ultraviolet-infrared
(UV-IR), correspondence in the amplitudes of perturbative string 
theories. The UV-IR correspondence manifests itself as modular 
invariance of the amplitudes of any closed string theory or,
equivalently, as the property of open-closed worldsheet duality 
in the amplitudes of a perturbative open and closed string theory
\cite{poltorus,polchinskibook,rg}. Unlike field theory, where 
finite temperature alters the infrared behavior of the theory,
leaving its renormalizability and short distance properties 
untouched--- at least at low temperatures, in perturbative string
theory it is simply not possible to alter the IR limit, without
simultaneously altering the UV behavior. 

\vskip 0.1in 
To understand why, consider 
exciting only the lowest-lying n=1 momentum mode in the zero
temperature vacuum. At low temperatures, the energy of the n=1
thermal vacuum is an apparently negligible shift from the 
energy of the zero temperature vacuum: $+ {{1}\over{\beta^2}}$,
$\beta$$>>$$1$. Nevertheless, modular invariance of the one-loop 
closed string amplitude has the consequence of introducing a 
Virasoro tower of states with string-scale masses contributing
to the intensive vacuum functional and, consequently, to the
Helmholtz free energy and the remaining thermodynamic potentials. 
The Virasoro tower arises from excitations of the infinity of 
allowed string oscillators in the n=1 thermal vacuum. Moreover, 
thermal self-duality of the heterotic 
theory or, in the case of type II, invariance under the thermal duality 
transformation mapping IIA to IIB, implies that, for any momentum mode 
in the spectrum, we must include a corresponding winding mode, w=1. 
Further, since the tower of oscillator excitations of the n=1 
and w=1 vacua have comparable mass to states in the Virasoro tower 
of the n=2, and w=2, thermal vacua, it would be inconsistent to leave 
these out. Iterating this argument, we see that even an {\em infinitesimal}
shift in temperature away from zero temperature in the closed string
vacuum, results in a vacuum functional, and free energy, with contributions
from an {\em infinite} sum over momentum and winding thermal vacua, 
each contributing a Virasoro tower of excited string states.  

\vskip 0.1in
In the Introduction, we have explained why the huge multiplicity of 
states in the string spectrum accounts for both the exponentially 
diverging density of states function in the UV limit and, as a
consequence of modular invariance, the absence of any signal
of this exponential degeneracy in the behavior of the free energy. 
That this is only true of a tachyon-free string theory is a 
consequence of the UV-IR relations: as was first noted in 
\cite{poltorus}, the decidedly IR phenomenon of a tachyonic 
state in the string spectrum can manifest itself as an apparent 
UV divergence in the free energy. This can be seen by a Poisson 
resummation of the modular invariant partition function, which 
switches the UV and IR asymptotics \cite{polchinskibook}.
In the superstring theories, the UV-IR consistency conditions 
have dramatic consequence. The absence of the zero temperature
tachyon in the supersymmetric spectrum, and the vanishing free
energy, are a consequence of the GSO projection: the cancellation
follows from the existence of modular invariant 
fermionic string partition functions which also respect the global
$SO(8)$ rotational symmetry of the transverse degrees of freedom
on the worldsheet in a Lorentz invariant 10D ground state. The
requirement of partition functions that are both invariant under 
the one-loop modular group, and which also respect the global $SO(8)$ 
symmetry, is remarkably restrictive. As a consequence, there are 
no viable alternatives to the supersymmetric combination of spin structures
other than those considered in \cite{rohm,aw}, and also in 
Sections 3 and 4. Thus, there are no \lq\lq small perturbations" 
of the supersymmetric GSO projection: modular invariance of the 
global $SO(8)$ invariant thermal partition functions destroys 
the tachyon cancellation mechanism in entirety, even for an 
infinitesimal variation in temperature away from zero temperature.

\vskip 0.1in
Our results are striking confirmation of the necessity to extend 
one's purvey of type II string theory to include gauge fields 
arising due to the presence of Dbrane or NSbrane fluxes. Whether 
this can be achieved at weak string coupling remains an interesting 
question. We note that it should be possible to use the extensive 
exploration of tachyon-free type 0 modular invariants in the recent 
literature, whether by the introduction of a Ramond-Ramond flux 
\cite{klebt}, or by suitable orbifold or orientation-reversal 
action on the type 0 partition function \cite{orient}. A Ramond-Ramond 
flux will shift the mass squared of the zero temperature tachyonic 
state to positive values. In this case, it should be straightforward 
to construct
a modular invariant type II vacuum functional which also transforms
correctly under thermal duality: 
namely, undergoing a Kosterlitz-Thouless continuous phase transition 
at $T$$=$$T_C$ which maps the thermal IIA string to the thermal 
IIB string, and with a tachyon-free spectrum at all temperatures 
above zero. An important issue is the clarification of a
natural choice of RR or NS fluxes in keeping with the 
interpretation of the tachyon-free finite temperature
ground state as a one-parameter fixed line describing 
a stable IIA or IIB thermal ensemble. This, of course, assumes a
resolution at weak string coupling.

\vskip 0.1in
In the event that the resolution of this puzzle involves a flow to 
strong type II string coupling, strong-weak coupling dualities in
the moduli space of String/M theory will be helpful. 
To begin with, recall that IIA/M theory orbifolds with Wilson 
line background for the 
Ramond-Ramond one-form potential have been widely studied in the 
duality literature \cite{cl}. There is a close correspondence
between the heterotic self-dual momentum lattice and the cohomology lattice
describing the moduli space of orbifolds of M/IIA theory 
compactified on quantum K3 
surfaces with Wilson line background for the RR one-form potential,
extensively explored in \cite{cl}. It is possible
that a Wilson line
background for the RR one-form potential gives a simple, and universal, 
resolution of the tachyonic instability in the thermal IIA string theory. 
Thermal duality can then be invoked to infer a similar 
resolution for the thermal IIB string. 

\vskip 0.1in
In this context, notice 
that the shift in momentum eigenvalue: 
$(n,w)$$\to$$(n+\half,w+\half)$, introduced 
in Eq.\ (\ref{eq:pfIItnsassd}) in order to 
obtain a stable thermal ensemble at temperatures of order
the string scale, results in a temperature-dependent 
shift in the vacuum energy of the $(n,w)$ thermal vacuum. This is
reminiscent of the 10D cosmological constant in the massive IIA
string \cite{dbrane}, the low energy effective limit of which is
Roman's massive 10D IIA supergravity \cite{romans}. 
Dimensional reduction of the massive IIA supergravity gives
a 9D massive supergravity which coincides with a Scherk-Schwarz 
dimensional reduction of the IIB theory \cite{ss,berg}.
Finally, while the massive IIA supergravity cannot be obtained from 
11D supergravity, or any covariant massive deformation thereof,
Hull \cite{hull} has pointed out that the massive IIA superstring 
nevertheless has a consistent M theory origin.
In previous work \cite{flux}, we have noted that the mass parameter
appears to be quantized in string soliton backgrounds of the 9D
massive supergravity with a consistent type II world-sheet description.
Nevertheless, in the absence of gauge fields,
the tachyonic instability in the thermal type II string
at a temperature of $O(\alpha^{\prime 1/2})$,
remains a puzzle. We remark that 
it is possible that the low energy description of the
putative stable and 
tachyon-free IIA or IIB thermal ensemble, of necessity, is
always a ground state of the massive 9D type II supergravity: 
a finite temperature vacuum
of the massive IIA string. As emphasized above, since IIA and IIB 
are related
by a thermal duality transformation in 9D, we may 
equivalently interpret this as
a thermal IIB vacuum. If true, this would give important insight into
M theory beyond its low energy 11D supergravity effective limit 
\cite{berg,hull,flux}.

\vskip 0.1in
It is also of interest to examine the full generality of the 
tachyonic instability at the string-scale temperature, $T_{\rm min}$,
found in the thermal IIA and IIB strings described in Section 3.1.
Consider compactifying this theory on an additional circle: 
namely, the IIA or IIB string compactified on the Euclidean 
space $R^8$$\times$$S^1$$\times$$S^1/Z_2$.
Upon compactification, the arguments given in section 3.1 for 
determining the supersymmetry 
breaking phases, and the general form of the
modular invariant combination of Jacobi theta functions in the
fermionic partition function, 
are unchanged. The mass formula gains
a new contribution from the nontrivial momentum and winding modes
in the compact spatial direction, $X^9$. Under a spatial
T$_9$-duality mapping IIA to IIB, we find that spatial 
IIA momentum modes are switched with spatial IIB winding modes, 
and vice versa, so that, from momentum conservation, the 
$n_9$$=$$w_9$$=$$0$ sector is retained in the physical state spectrum. 
This sector of the theory has thermal tachyons obeying the mass
formula given in Eq.\ (\ref{eq:massII}). Thus, by the same reasoning
as given in section 3.1, the tachyonic thermal modes of the 
thermal flat space type II string are retained in the zero spatial 
momentum sector of the compactified thermal type II string.

\vskip 0.1in
Notice that further compactification, or 
orbifolding, of additional spatial coordinates
has no impact whatsoever on the thermal tachyonic instability in the
untwisted zero spatial momentum sector of the compactified
thermal type II string.
We comment that taking the large radius limit of the Euclidean space, 
$R^m$$\times$$T^{9-m}$$\times$$S^1/Z_2$, or any $Z_N$ orbifold
thereof, where the $Z_N$'s act as spatial rotations and the orbifold 
singularities are appropriately repaired in the noncompact limit,
will give an ALE space: an asymptotically locally Euclidean space 
of the general form $R^k/\Gamma$. 
In our example, taking the large radius limit for all ten coordinates
describes the asymptotic approach to a noncompact type II
vacuum in the vicinity of the zero temperature supersymmetric 
vacuum. However, in the absence of a timelike Wilson line, 
the thermal tachyonic instability expected at the string-scale minimum 
temperature would appear to be an obstruction to obtaining a 
noncompact ALE space in the timelike direction. This appears to be
a counter-example to the cases discussed in \cite{aps}.

\vskip 0.1in
The broader question of whether the formation of a condensate in a
tachyonic closed string ground state can drive a phase transition 
to a tachyon-free vacuum---
non-supersymmetric, or supersymmetric, is of considerable interest 
both in the context of finite temperature instabilities, but also
for a deeper understanding of the configuration space of String/M theory. 
This question has been addressed in recent work \cite{aps} 
in the context of the conjectured IIA-type 0 dualities 
\cite{tseytlin,klebt,gab}. The $O(\alpha^{\prime})$ analysis of 
the fate of closed string tachyons in nonsupersymmetric ALE spacetimes 
in \cite{aps} suggests that a chain of such phase transitions 
will always terminate in a stable 
supersymmetric ALE spacetime. We note that a more detailed 
analysis of this issue using worldsheet methods, and in compact 
flat spacetime geometries, may be viable in the heterotic string
theory. 
Recall that tachyonic vertex operators correspond to relevant perturbations
of the worldsheet superconformal field theory and, from Zamolodchikov's
c theorem, the RG flow is towards a fixed point of lower 
matter central charge. In the manifestly Weyl invariant framework 
of string theory, this implies nontrivial dynamics for the Liouville
field along the flow to the new fixed point.\footnote{As an aside, 
note that it is perfectly consistent with the c-theorem that the 
number of noncompact spacetime dimensions at the new fixed point
is lower: a dynamical mechanism for choosing a stable ground state
with noncompact spacetime dimension $D$$<$$10$, similar in spirit to 
the dynamical approach pursued in the nonperturbative matrix 
formulation \cite{ikkt,bfss}.}

\vskip 0.1in
Ordinarily, this would be considered an incalculable problem because 
of the famed ${\hat{c}}$$=$$1$ barrier \cite{ddk} to an 
analytic treatment of the Liouville superconformal field theory.
But in the context of the finite temperature induced tachyonic 
instability it may be possible to address relevant flows:
the ($X^0_E$,$\phi_L$) dynamics is a 2D superconformal field theory 
coupled to 2D supergravity, which fits nicely within the realm of 
solvable 2D field theories \cite{ddk,jp}. We should point out that the
modular invariant 10D nonsupersymmetric and tachyonic heterotic 
partition functions have been classified, both in an 
analysis of abelian orbifolds by Dixon and Harvey \cite{dh}, and 
also in the free fermionic spin structure construction by 
Kawai, Lewellen, and Tye \cite{klt}. Thus, all of the potential
tachyonic noncompact limits of the stable and tachyon-free
finite temperature heterotic ground state are known. It is natural 
to conjecture that any unstable 10D heterotic vacuum would 
of necessity flow to the unique tachyon-free 9D 
nonsupersymmetric heterotic string ground state, and with 
gauge group $SO(16)$$\times$$SO(16)$. 
Furthermore, we should note that there already exists a nonperturbative
framework for 2D bosonic string theory in the form of the c=1 
matrix model \cite{review}, with extensive studies of tachyon 
dynamics, although the ${\hat{c}}$$=$$1$ super-matrix model remains
puzzling in many respects.
Nevertheless, we find it intriguing that an exactly 
solvable 2D theory could provide insight into the 
nonperturbative finite coupling 
behavior of realistic supersymmetric heterotic
string theories at finite temperature.

\vskip 0.1in
In \cite{us}, 
motivated by a knowledge of the 
phase structure of the simplest lattice field theories, 
and the combined 
conjectural and numerical evidence for the two-parameter 
$(g_{YM},\beta)$ phase diagram of continuum Yang-Mills theory 
at finite temperature and finite gauge coupling, 
we noted that it is natural to ask whether analogous
questions can be addressed in String/M theory---- 
even in limited regions of the moduli space. 
We emphasize that our goal here is not further elucidation of finite 
temperature gauge theory at strong coupling--- although this would be an 
added plus. Instead, the general theme of the work in 
\cite{flux,us,type1,rg} has been to invoke analogy 
with gauge theory in order to motivate world-sheet computations 
probing sub-string scale distances which might give insight 
into nonperturbative aspects of String/M theory, 
possibly even at strong coupling. The spontaneous breaking 
of thermal duality by thermal Dbranes in the type I$^{\prime}$
string, and a worldsheet computation
of the pair correlator of timelike Wilson loops
with fixed loop separation $r$$<$$\alpha^{\prime 1/2}$ 
in the tachyon-free thermal type I$^{\prime}$ ground state, are
explored at length in \cite{us,type1}. As in ordinary 
gauge theory, the correlator serves as an order parameter 
for a continuous phase transition in the thermal type I$^{\prime}$ string
at a string-scale transition temperature computed in \cite{us,type1}:
above the transition temperature, we find that
the order parameter vanishes asymptotically
in the $\beta$$\to$$0$ high temperature limit, signaling 
deconfinement.

\vskip 0.1in
We conclude with the encouraging observation that our 
results are evidence for the limited validity of strong-weak
coupling duality conjectures in the absence of spacetime 
supersymmetry \cite{cl,vw,dienes,orb,harvey,gab}: 
an infinitesimal background temperature 
cannot change the nature of the 
weak-strong coupling dualities holding in 
the vicinity of a non-pathological weakly-coupled 
supersymmetric ground state.
Thus, the strong-weak coupling duality relations 
valid for the supersymmetric dual string pair, 
are expected to hold for the tachyon-free 
finite temperature dual string pair, 
at least at infinitesimal temperature.
Of necessity, such dual string pairs will 
exhibit broken supersymmetry with inverse radius, $1/\beta$, 
playing the role of a small susy-breaking control parameter. 

\vspace{0.2in}
\noindent{\bf Acknowledgements:}
Many of our results can be anticipated from the analysis of the 
duality transition and thermodynamics of a gas of free closed 
bosonic strings given in \cite{bosonic} (hep-th/0105110), which 
should be read in conjunction with this paper. I would like to 
thank K. Dienes, A. Dhar, J. Distler, M. Fukuma, J. Harvey, H. Kawai, D. 
Kutasov, M. Peskin, E. Silverstein, B. Sundborg, and E. Witten, 
for raising interesting questions on this work. This research is 
supported in part by the award of grant
NSF-PHY-9722394 by the National Science Foundation under the 
auspices of the Career program.

\vskip 0.5in
\noindent{\bf Note Added (Sep 2005):} Many of the points made in this
paper are either extraneous, or incorrect in the details, although the broad
conclusions summarized in the abstract
 stand. Namely, that there is no self-consistent type II 
superstring ensemble, in the absence of a Yang-Mills gauge sector.
And that both heterotic and type I theory have equilibrium canonical
ensembles free of thermal tachyons; a crucial role is played by the
temperature dependent Wilson line wrapping Euclidean time. I refer 
the reader to hep-th/0506143.


\vspace{0.3in}
\begin{thebibliography}{99}
\bibitem{kirzhlinde} 
S. Coleman and E. Weinberg, 
{\em Radiative Corrections as the Origin of Spontaneous
Symmetry Breaking},  Phys.\ Rev.\ {\bf D7} (1973) 1888.
R.\ Jackiw, {\em } Phys.\ Rev.\ {\bf D9} (1974) 1686.
M. Peskin and D. Schroeder, {\em An Introduction to Quantum Field 
Theory}, Addison-Wesley (1995).  
A. Linde, {\em Phase Transitions in Gauge Theories and Cosmology},
Rept.\ Prog.\ Phys.\ {\bf 42} (1979) 389.
\bibitem{gpy} D. Gross, R. Pisarski, and L. Yaffe,
{\em QCD and Instantons at Finite Temperature},
Rev.\ Mod.\ Phys.\ {\bf 53} (1981) 43.
J. Kapusta, {\em Finite Temperature Field Theory}, Cambridge
(1989). A. Das, {\em Finite Temperature Field Theory}, World Scientific
(1997).
\bibitem{bernard}C. Bernard, Phys.\ Rev.\ {\bf D9} (1974)
3312.
\bibitem{dinesei} M. Dine and N. Seiberg, {\em Is the Superstring
Weakly Coupled?}, Phys.\ Lett.\ {\bf B162} 299 (1985).
\bibitem{aw} J. Atick and E. Witten,
{\em The Hagedorn Transition and the Number of Degrees of Freedom in
String Theory},
Nucl.\ Phys.\ {\bf B310} (1988) 291.
\bibitem{us} S. Chaudhuri, {\em Deconfinement and the Hagedorn Transition in
Thermal String Theory}, Phys.\ Rev.\ Lett.\ {\bf 86}, Issue 10, 1943 (2001),
hep-th/0008131.
\bibitem{rohm}R. Rohm, {\em Spontaneous Supersymmetry Breaking
in Supersymmetric String Theories}, Nucl.\ Phys.\ {\bf B237} (1984) 553.
\bibitem{bowick}
M. Bowick and L.C.R. Wijewardhana, {\em Superstrings at High Temperature},
Phys.\ Rev.\ Lett.\ {\bf 54} 2485 (1985).
\bibitem{poltorus}J. Polchinski, 
{\em Evaluation of the One Loop String Path Integral},
Comm. Math. Phys. {\bf 104} (1986) 37.
\bibitem{mcroth}B. Maclain and B. Roth, Comm.\ Math.\ Phys.\
{\bf 111} (1987) 1184.
\bibitem{kosterlitz} J. M. Kosterlitz, 
J.\ Phys.\ {\bf C7} (1974) 1046. J. Cardy, {\em Scaling 
and Renormalization in Statistical Physics}, Chap.\ 6 (Cambridge) 1996.
C. Itzykson and J. Drouffe,
{\em Statistical Field Theory}, Vol.\ I, sec.\ 4.2.
\bibitem{polchinskibook} J. Polchinski, {\it String Theory},
Vol.\ II (Cambridge) 1998. See, in particular, sections 10.7
and 11.6. 
\bibitem{polyakov} A. M. Polyakov,
{\em Quantum Geometry of Fermionic Strings},
Phys.\ Lett.\ {\bf B103} 211 (1980).
\bibitem{rg} S. Chaudhuri, {\em The Power of World Sheets:
Applications and Prospects}, talk given at Tohwa International
Symposium on String Theory, Jul 3-7 (2001), Fukuoka, Japan. 
PSU-TH-245.
Online at: http://www.tohwa-u.ac.jp/~tada/symposium/speakers.html.
S. Chaudhuri and E. Novak, {\em Effective String Tension and
Renormalizability: String Theory in a Noncommutative Space},
JHEP {\bf 08} (2000) 027.
\bibitem{gv} P. Ginsparg and C. Vafa,
{\em Toroidal Compactifications of Nonsupersymmetric
Heterotic Strings}, Nucl.\ Phys.\ {\bf B289} (1987) 414.
\bibitem{itoyama}
H. Itoyama and T. Taylor, 
{\em Supersymmetry Restoration in the Compactified 
$O(16)$$\times$$O(16)$ Heterotic String Theory},
Phys.\ Lett.\ {\bf B186} (1987) 129.
\bibitem{hagedorn} R. Hagedorn, Nuovo Cim.\ Suppl.\ 3 (1965) 147.
K. Huang and S. Weinberg,
Phys.\ Rev.\ Lett.\ {\bf 25} (1970) 895.
S. Fubini and G.  Veneziano, Nuovo Cim.\ {\bf 64A} (1969) 1640.
R. Carlitz, Phys. Rev. {\bf D5} 3231 (1972).
S. Frautschi, Phys. Rev. {\bf D3} 2821 (1971).
N. Deo, S. Jain, and C.-I. Tan, 
{\em String Statistical Mechanics above Hagedorn Energies},
Phys. Rev. {\bf D36} (1987) 1184.
M. Bowick and S. Giddings, {\em High Temperature Strings},
Nucl.\ Phys.\ {\bf B325} (1989) 631.
D. Lowe and L. Thorlacius, {\em Hot String Soup},
Phys.\ Rev.\ {\bf D51} (1995) 665.
\bibitem{tan} 
K. Kikkawa and M. Yamasaki, {\em
Casimir Effects in Superstring Theories},
Phys.\ Lett.\ {\bf B149} (1984) 357.
N. Sakai and I. Senda,
{\em Vacuum Energies of String Compactified on Torus},
Prog.\ Theor. Phys. 75 (1986) 692.
B. Sathiapalan, {\em 
Duality in Statistical Mechanics and String Theory},
Phys.\ Rev.\ Lett.\ {\bf 58} (1987) 414.
Ya. I. Kogan, {\em Vortices on the World-Sheet and String's
Critical Dynamics}, JETP Lett.\ {\bf 45} (1987) 709. 
K. O'Brien and Chung-I Tan,
{\em Modular Invariance of Thermopartition Function and Global
Phase Structure of Heterotic String},
Phys.\ Rev.\ {\bf D36} (1987) 1184.
V.P. Nair, A. Shapere, A. Strominger,
and F. Wilczek, {\em Compactification of the Twisted Heterotic String},
Nucl.\ Phys.\ {\bf B287} (1987) 402.
R. Brandenberger and C. Vafa,
{\em Superstrings in the Early Universe}, 
Nucl.\ Phys.\ {\bf B316} (1989) 391.
\bibitem{bosonic}S. Chaudhuri, {\em Finite Temperature Closed
Bosonic Strings: Thermal Duality and the KT Transition},
PSU-TH-241, hep-th/0105110.
\bibitem{dgh} L. Dixon, P. Ginsparg, and J. Harvey,
{\em Central Charge ${\hat C}$$=$$1$ Superconformal Field
Theory}, Nucl.\ Phys.\ {\bf B306} (1988) 470.
See, also, {\em Beauty and the Beast: Superconformal Symmetry in
a Monster Module}, Comm.\ Math.\ Phys.\ {\bf 119} (1988) 285.
\bibitem{nsw} K.S. Narain, M. Sarmadi, and E. Witten,
{\em A Note on Toroidal Compactification of Heterotic String
Theory}, Nucl.\ Phys.\ {\bf B279} (1987) 369.
\bibitem{ginine} P. Ginsparg, {\em Comment on Toroidal 
Compactification of Heterotic Superstrings},
Phys.\ Rev.\ {\bf D35} 648 (1987).
\bibitem{sw}N. Seiberg and E. Witten, {\em Spin Structures in String
Theory}, Nucl.\ Phys.\ {\bf B276} (1986) 272.
\bibitem{agmv}L. Alvarez-Gaume, P. Ginsparg, G. Moore, and C. Vafa,
{\em An $O(16)$$\times$$O(16)$ Heterotic String},
Phys.\ Lett.\ {\bf B171} (1986) 155. 
\bibitem{dh}L. Dixon
and J. Harvey, {\em String Theories in Ten Dimensions
Without Spacetime Supersymmetry},
Nucl.\ Phys.\ {\bf B274} 93 (1986). 
\bibitem{klt}H. Kawai, D. Lewellen, and S.-H. H. Tye, 
{\em Classification of Closed Fermionic String Models},
Phys.\ Rev.\ {\bf D34} (1986) 3794.
\bibitem{ckt} S. Chaudhuri, H. Kawai, and S.-H.H. Tye,
{\em Path Integral Formulation of Strings}, Phys.\ Rev.\ {\bf D36} 1148
(1987). See section 3.
\bibitem{wein}E.\ Witten, Princeton preprints (1985).
See discussion in S.\ Weinberg, {\em Coupling Constants and
Vertex Functions in String Theories},  
Phys.\ Lett.\ {\bf 156B} 309.
\bibitem{zam}A.\ B.\ Zamolodchikov, {\em Irreversibility of the Flux of the
RG in a 2D field theory}, JETP Letters, {\bf 43}, 730.
\bibitem{aps}A.\ Adams, J.\ Polchinski, E.\ Silverstein,
{\em Don't Panic! Closed String Tachyons in ALE Spacetimes},
hep-th/0108075. 
\bibitem{gyp} D. Gross, L. Yaffe, and M. Perry,
{\em Instability of Flat Space at Finite Temperature},
Phys.\ Rev.\ {\bf D25} (1982) 330.
\bibitem{tseytlin}F. Dowker, J.\ P.\ Gauntlett, D.\ A.\ Kastor,
and J. Traschen, {\em Pair Creation of Dilaton Black Holes},
Phys.\ Rev.\ {\bf D49} (1994) 2909.
J.\ G.\ Russo and A.\ A.\ Tseytlin, {\em Constant Magnetic Field
in Closed String Theory: An Exactly Solvable Model}, Nucl.\ Phys.\
{\bf B448} (1995) 293.
A.\ A.\ Tseytlin, {\em Melvin Solution in
String Theory}, Phys.\ Lett.\ {\bf B346} (1995) 55.
A.\ A.\ Tseytlin, Talk given 
at Tohwa International Symposium on String Theory,
Fukuoka, Japan. Jul 3-7, 2001. 
Online at: http://www.tohwa-u.ac.jp/~tada/symposium/speakers.html
\bibitem{gab}O.\ Bergman and M.\ Gaberdiel,
{\em Dualities of Type 0 Strings}, JHEP 9907, 022 (1999).
Y.\ Imamura, {\em Branes in Type 0/Type II Duality},
hep-th/9906090.
M. Gaberdiel and A.\ Strominger, {\em Fluxbranes in 
String Theory}, JHEP 0106, 035 (2001).
\bibitem{klebt} I.\ R.\ Klebanov and A.\ A.\ Tseytlin, {\em Dbranes and 
Dual Gauge Theories in Type 0 String Theory}, Nucl.\ Phys.\
{\bf B546} (1999) 155.
\bibitem{orient}C.\ Angelatonj, {\em Non-Tachyonic Open Descendents
of the 0B String Theory}, hep-th/9810214. R.\
Blumenhagen, A.\ Font, and D.\ Lust, {\em tachyon-free Orientifolds
of Type 0B Strings in Various Dimensions}, hep-th/9904069.
\bibitem{chl}S. Chaudhuri, G. Hockney, and J. Lykken,
{\em Maximally Supersymmetric String Theories in D<10},
Phys.\ Rev.\ Lett.\  {\bf 75} (1995) 7168. 
S. Chaudhuri and J. Polchinski, {\em Moduli Space of the CHL Orbifold},
Phys.\ Rev.\ {\bf D52} (1995) 7168.
\bibitem{ddk}
V.\ G.\ Knizhnik, A.\ M.\ Polyakov, and
A.\ B.\ Zamolodchikov,
Mod.\ Phys.\ Lett.\ {\bf A3} (1988) 819.
J.\ Distler and H.\ Kawai, Nucl.\ Phys.\ {\bf B321} (1989) 509.
F.\ David, Mod.\ Phys.\ Lett.\ {\bf A3} (1988) 1651.
J. Distler, H. Kawai, and Z.\ Hloussek, {\em SuperLiouville
Theory as a Two Dimensional Superconformal Supergravity Theory},
Int.\ J.\ Mod.\ Phys.\ {\bf A5} 391 (1990). 
\bibitem{dk}
D.\ Kastor, E.\ Martinec, and S.\ Shenker,
{\em RG Flow in N=1 Discrete Series}, Nucl.\ Phys.\ {\bf B316}
(1989) 590. 
P.\ de Francesco, J.\ Distler, D.\ Kutasov,
{\em Superdiscrete Series Coupled to 2D Supergravity},
Mod.\ Phys.\ Lett.\ {\bf A5} 2135 (1990).
\bibitem{jp}J.\ Polchinski, {\em A Two-Dimensional Model for
Quantum Gravity}, Nucl.\ Phys.\ {\bf B324} (1989) 123.
A.\ M.\ Polyakov, {\em A Few Projects in String Theory}, hep-th/9304146.
See section 9 on scale dependence in quantum gravity. I. Klebanov,
I.\ I.\ Kogan, and A.\ M.\ Polyakov, {\em Gravitational Dressing
of the Renormalization Group}, Phy.\ Rev.\ Lett.\
{\bf 71} (1993) 3243.
\bibitem{ikkt}N.\ Ishibashi, H.\ Kawai, Y.\ Kitazawa, and
A.\ Tsuchiya, {\em A Large N Reduced Model as Superstring},
Nucl.\ Phys.\ {\bf B498} (1997) 467. H.\ Aoki,
S.\ Iso, H.\ Kawai, Y.\ Kitazawa, and T.\ Tada,
{\em Spacetime Structures from IIB Matrix Model},
prog.\ Theor.\ Phys.\ {\bf 99} (1998) 713. H.\ Kawai,
{\em Constructing New Types of Matrix Models},
Talk given at Tohwa International Symposium on
String Theory, Fukuoka, Japan. Jul 3-7, 2001.  
Online at: http://www.tohwa-u.ac.jp/~tada/symposium/speakers.html
\bibitem{bfss}
T. Banks, W. Fischler, S. Shenker, and L. Susskind,
{\em M Theory as a Matrix Model}, Phys.\ Rev.\ {\bf D55} (1997) 5112.
\bibitem{review}P.\ Ginsparg and G.\ Moore,
{\em Lectures on 2D gravity and 2D string theory}, TASI lectures,
hep-th/9304011.
\bibitem{cl}J.\ A.\ Schwarz and A.\ Sen, {\em The Type IIA Dual of the
Six-Dimensional CHL Compactification}, Phys.\ Lett.\ {\bf B357} 323 (1995).
S.\ Chaudhuri and D.\ Lowe, {\em Type IIA-Heterotic Duals with
Maximal Supersymmetry}, Nucl.\ Phys.\ {\bf B459} (1996) 113.
{\em Monstrous String-String Duality}, Nucl.\ Phys.\ {\bf B469},
(1996) 21. 
\bibitem{dbrane}J.\ Polchinski, {\em Dirichlet-branes and 
Ramond-Ramond Charge}, Phys.\ Rev.\ Lett.\ {\bf 75} (1995) 4724.
\bibitem{romans}L.\ Romans, Phys.\ Lett.\ {\bf B169} (1986) 374.
\bibitem{ss}J. Scherk and J. Schwarz, {\em Spontaneous Breaking of Supersymmetry
through Dimensional Reduction}, Phys.\ Lett.\ {\bf B82} (1979). 
\bibitem{berg}E.\ Bergshoeff, M.\ de Roo, M.\ B.\ Green, 
G.\ Papadopoulos, and P.\ Townsend, {\em Duality of Type II 7-branes and 
8-branes}, Nucl.\ Phys.\ {\bf B470} (1996) 113.
\bibitem{hull}C.\ Hull, {\em Massive String Theories from M theory
and F theory}, JHEP 9811 (1998) 027. 
\bibitem{flux}S.\ Chaudhuri, {\em Confinement and the Short
Type I$^{\prime}$ Flux Tube}, Nucl.\ Phys.\ {\bf B591} (2000)
243.
\bibitem{type1} S. Chaudhuri, {\em Spontaneously Broken Thermal 
Duality: Phase Transitions in the Heterotic-Type I String}, PSU-TH-244.
\bibitem{vw} S. Ferrara, J.\ Harvey, 
A.\ Strominger, and C.\ Vafa, {\em Second Quantized Mirror
Symmetry}, hep-th/9505162.
C.\ Vafa and E.\ Witten, {\em Dual String Pairs with
N=1 and N=2 Supersymmetry in D=4}, hep-th/9507050. 
J.\ Harvey, A.\ Strominger, and D.\ Lowe, {\em N=1
String Duality}, hep-th/9507168. P.\ Aspinwall, {\em An N=2
Pair and a Phase Transition}, hep-th/9510142.
H. Gao, {\em More Dual
String Pairs from Orbifolding}, hep-th/9512060. K.\
Dasgupta and S.\ Mukhi, {\em Orbifolds of M Theory}, Nucl.\
Phys.\ {\bf B465} (1996) 399.
\bibitem{dienes} K. Dienes and J. Blum, {\em Duality without Supersymmetry:
The Case of the $SO(16)$$\times$$SO(16)$ String}, Phys.\ Lett.\
{\bf B414} (1997) 260. {\em Strong-Weak Coupling Duality
Relations for Non-supersymmetric String Theories},
Nucl.\ Phys.\ {\bf B516} (1998) 83.
\bibitem{orb} S. Kachru, J. Kumar, and E. Silverstein, {\em Vacuum Energy 
Cancellations in a Nonsupersymmetric 
String}, Phys.\ Rev.\ {\bf D59} (1999) 106004.
G. Shiu and S.-H. H. Tye, {\em Bose-Fermi Degeneracy and Duality 
in Nonsupersymmetric Strings}, Nucl.\ Phys.\ {\bf B542} (1999) 45.
\bibitem{harvey} J. Harvey, {\em String Duality and 
Non-supersymmetric String Theories}, Phys.\ Rev.\ {\bf D59} (1998) 026002. 
\end{thebibliography}
\end{document}